\documentclass[journal]{IEEEtran}
\usepackage[LGR,T1]{fontenc}
\usepackage{array}
\usepackage{booktabs}
\usepackage{enumitem}
\setlist[enumerate]{label=\roman*-} 
\usepackage{hyperref}
\usepackage{pbox}
\usepackage{amsmath}
\usepackage{balance}
\usepackage[byname]{smartref}
\usepackage{amsmath,amssymb,amsfonts} 
\usepackage{color}                    % For creating coloured text and background
\usepackage{comment}
\usepackage{mathtools}
\usepackage{commath}
\usepackage{graphics,graphicx}
\usepackage{epstopdf}
\usepackage[font={small,it}]{caption}
\usepackage{placeins}
\usepackage{tikz}
\usepackage{bchart}
\usepackage{enumitem}
\usepackage{adjustbox}
\usepackage{pstricks,pst-node,pst-tree}
\usepackage{pst-plot}

\makeatletter

\ProvideTextCommand{\~}{LGR}[1]{\char126#1}

\providecommand{\tabularnewline}{\\}

\makeatother

\begin{document}

\title{Multi-Stage Optimized Machine Learning Framework for Network Intrusion Detection}

\author{\IEEEauthorblockN{MohammadNoor Injadat\IEEEauthorrefmark{1}, Abdallah Moubayed\IEEEauthorrefmark{1},
Ali Bou Nassif\IEEEauthorrefmark{2}\IEEEauthorrefmark{1},
Abdallah Shami\IEEEauthorrefmark{1}}\\
\IEEEauthorblockA{\IEEEauthorrefmark{1}Department of Electrical and Computer Engineering,
University of Western Ontario, London, ON, Canada \\
Email: \{minjadat, amoubaye, abdallah.shami\}@uwo.ca}\\
\IEEEauthorblockA{\IEEEauthorrefmark{2}Department of Computer Engineering,
University of Sharjah, Sharjah, UAE\\
Email: anassif@sharjah.ac.ae}}

\maketitle
\begin{abstract}
Cyber-security garnered significant attention due to the increased dependency of individuals and organizations on the Internet and their concern about the security and privacy of their online activities. Several previous machine learning (ML)-based network intrusion detection systems (NIDSs) have been developed to protect against malicious online behavior. This paper proposes a novel multi-stage optimized ML-based NIDS framework that reduces computational complexity while maintaining its detection performance. This work studies the impact of oversampling techniques on the models' training sample size and determines the minimal suitable training sample size. Furthermore, it compares between two feature selection techniques, information gain and correlation-based, and explores their effect on detection performance and time complexity. Moreover, different ML hyper-parameter optimization techniques are investigated to enhance the NIDS's performance. The performance of the proposed framework is evaluated using two recent intrusion detection datasets, the CICIDS 2017 and the UNSW-NB 2015 datasets. Experimental results show that the proposed model significantly reduces the required training sample size (up to 74\%) and feature set size (up to 50\%). Moreover, the model performance is enhanced with hyper-parameter optimization with detection accuracies over 99\% for both datasets, outperforming recent literature works by 1-2\% higher accuracy and 1-2\% lower false alarm rate.
\end{abstract}


\IEEEpeerreviewmaketitle{}

\section{Introduction}\label{intro}
\indent The Internet has become an essential aspect of daily life with individuals and organizations depending on it to facilitate communication, conduct business, and store information \cite{bib1,sb}. This dependence is coupled with these individuals and organizations' concern about the security and privacy of their online activities \cite{bib25}. Accordingly, the area of cyber-security has garnered significant attention from both the industry and academia. To that end, more resources are being deployed and allocated to protect modern Internet-based networks from potential attacks or anomalous activities. Several protection mechanisms have been proposed such as firewalls, user authentication, and the deployment of antivirus and malware programs as a first line of defense \cite{bib2}. However, these mechanisms have not been able to completely protect the organizations' networks, particularly with contemporary attacks \cite{bib3}.\\
\indent Typically, network intrusion detection systems (NIDSs) can be divided into two main categories: signature-based detection systems (misused detection) and anomaly-based detection systems \cite{bib4}. Signature-based detection systems base their detection on the observation of pre-defined attack patterns. Thus, they have proven to be effective for attacks with well-known signatures and patterns. However, these systems are vulnerable against new attacks due to their inability to detect new attacks by learning from previous observations \cite{bib5}. In contrast, anomaly-based detection systems base their detection on the observation of any behavior or pattern that deviates from what is considered to be normal. Therefore, these systems can detect unknown attacks or intrusions based on the built models that characterize normal behavior \cite{bib6}.\\ 
\indent Despite the continuous improvements in NIDS performance, there is still room for further improvement. This is particularly evident given the high volume of generated network traffic data, continuously evolving environments, vast amounts of features collected that form the training datasets (high dimensional datasets), and the need for real-time intrusion detection \cite{bib7}. For example, having redundant or irrelevant features can have a negative impact on the detection capabilities of NIDSs as it slows down the model training process. Therefore, it is important to choose the most suitable subset of features and optimize the parameters of the machine learning (ML)-based detection models to enhance their performance \cite{bib8}.\\
\indent This paper extends our previous work in \cite{sc} by proposing a novel multi-stage optimized ML-based NIDS framework that reduces the computational complexity while maintaining its detection performance. To that end, this work first studies the impact of oversampling techniques on the models' training sample size and determines the minimum suitable training size for effective intrusion detection. Furthermore, it compares between two different feature selection techniques, namely information gain and correlation-based feature selection, and explores their effect on the models' detection performance and time complexity. Moreover, different ML hyper-parameter optimization techniques are investigated to enhance the NIDS's performance and ensure its effectiveness and robustness.\\ %The main differences between this work and our previous work in \cite{sc} are as follows: 
%\begin{itemize}
%	\item This work first studies the suitable training sample size by investigating the training accuracy and cross-validation accuracy illustrated using the learning curve.
%	\item This work addresses the class imbalance problem by using an oversampling technique. This was not considered in the conference paper.
%	\item This work investigated the use of feature selection and studied its impact on the training sample size, feature set size, and consequently on the overall training complexity.
%	\item This work compares the performance of five different optimization techniques and studies their impact on the overall detection performance of the proposed ML-models. In the conference paper, only one optimization technique was considered. 
%	\item This work considered more recent datasets, namely the CICIDS 2017 (which is an extension of the previously used ISCX 2012 dataset) and the UNSW-NB 2015 dataset.
%\end{itemize}
\indent To evaluate the performance of the proposed optimized ML-based NIDS framework, two recent state-of-the-art intrusion detection datasets are used, namely the CICIDS 2017 dataset \cite{cicids17} (which is the updated version of the ISCX 2012 dataset \cite{bib9} used in our previous work \cite{sc}) and the UNSW-NB 2015 dataset \cite{unsw15}. The performance evaluation is conducted using various evaluation metrics such as accuracy (acc), precision, recall, and false alarm rate (FAR).\\ 
\indent The remainder of this paper is organized as follows: Section \ref{related} briefly summarizes some of the previous literature works that focused on this research problem and presents its limitations. Section \ref{contributions} summarizes the contributions of this work. Section \ref{mathematical_background} discusses the  theoretical mathematical background of the different deployed techniques. Section \ref{proposed_framework} presents the proposed multi-stage optimized ML-based NIDS framework. Section \ref{datasets} describes the two datasets under consideration in more details. Section \ref{performance} presents and discusses the experimental results obtained. Finally, Section \ref{conc} concludes the paper and proposes potential future research endeavors. 
\section{Related Work and Limitations}\label{related}
\subsection{Related Work}
\indent ML classification techniques have been proposed as part of various network attack detection frameworks and  other applications using different classification models such as Support Vector Machines (SVM) \cite{bib10}, Decision Trees \cite{bib11}, KNN \cite{bib12}, Artificial Neural Networks (ANN) \cite{ann1,s11}, and Naive Bayes \cite{bib13} as illustrated in \cite{bib1}. One such application is the DNS typo-squatting attack detection framework presented in \cite{se,se2}. Also, ML techniques have been proposed to detect zero-day attacks as illustrated by the probabilistic Bayesian network model presented in \cite{forensics1}. Comparatively, hybrid ML-fuzzy logic-based system that focuses on distributed denial of service (DDoS) attack detection has been proposed in \cite{TNSM2}. These ML classification techniques have also been proposed for bot net detection \cite{TNSM1} as well as for mobile phone malware detection \cite{forensics2}. \\
\indent Similarly, several previous works focused on the use of ML classification techniques for network intrusion detection. For example, Salo \textit{et al.} conducted a literature survey and identified 19 different data mining techniques commonly used for intrusion detection \cite{sf,sf1}. The result of this review highlighted the need for more ML-based research to address real-time IDSs. The authors then proposed an ensemble feature selection and an anomaly detection method for network intrusion detection \cite{sg}. In contrast, Li \textit{et al.} proposed a decision tree (DT)-based IDS model for autonomous and connected vehicles \cite{sd}. The goal of the IDS is to detect both intra-vehicle and external vehicle network attacks \cite{sd}.\\
\indent In a similar fashion, several previous research works proposed the use of various optimization techniques to enhance the performance of their NIDSs. For example, Chung and Wahid proposed a hybrid approach that included feature selection and classification with simplified swarm optimization (SSO) in addition to using weighted local search (WLS) to further enhance its performance \cite{bib14}. Similarly, Kuang \textit{et al.} presented a hybrid GA-SVM model associated with kernel principal component analysis (KPCA) to improve the performance \cite{bib15}. Comparatively, Zhang \textit{et al.} combined  misuse and anomaly detection using RF \cite{bib17}. In contrast, our previous work in \cite{sc} proposed a Bayesian optimization model to hyper-tune the parameters of different supervised ML algorithms for anomaly-based IDSs \cite{sc}. %More specifically, they tune the parameters of SVM, Random Forest (RF), and K-nearest neighbors (KNN) algorithms.
\subsection{Limitations of Related Work}
\indent Despite the many previous works in the literature that focused on the intrusion detection problem, the previously proposed models suffer from various shortcomings. For example, many of these works do not focus on the class imbalance issue often encountered in intrusion detection datasets. Also, the training sample size is often selected randomly rather than using a systematic approach. They are also limited by the use of outdated datasets such as NLS KDD99. Additionally, the results reported are usually only done using one dataset rather than being validated using multiple datasets. Few works also considered the hyper-parameter optimization using different techniques and used only one method instead. Also, only some research works studied the time complexity of their proposed framework, a metric that is often overlooked.  
\section{Research Contributions}\label{contributions}
\indent The main contributions and differences between this work and our previous work in \cite{sc} can be summarized as follows:   
\begin{itemize}
	\item Propose a novel multi-stage optimized ML-based NIDS framework that reduces computational complexity and enhances detection accuracy.
	\item Study the impact of oversampling techniques and determine the minimum suitable training sample size for effective intrusion detection.
	\item Explore the impact of different feature selection techniques on the NIDS detection performance and time (training and testing) complexity.
	\item Propose and investigate different ML hyper-parameter optimization techniques and their corresponding enhancement of the NIDS detection performance.
	\item Evaluate the performance of the optimized ML-based NIDS framework using two recent state-of-the-art datasets, namely the CICIDS 2017 dataset \cite{cicids17} and the UNSW-NB 2015 dataset \cite{unsw15}.
	\item Compare the performance of the proposed framework with recent works from the literature and illustrate the improvement of detection accuracy, reduction of FAR, and a reduction of both the training sample size and feature set size. 
\end{itemize}
To the best of our knowledge, no previous work proposed such a multi-stage optimized ML-based NIDS framework and evaluated it using these datasets.
\section{Background and Preliminaries}\label{mathematical_background}
\indent As mentioned earlier, this paper proposes a multi-stage optimized ML-based NIDS framework that reduces computational complexity while maintaining its detection performance. Multiple techniques are deployed at different stages for this to be implemented. An overview of the used techniques is given in what follows.
%\vspace{-0.7cm}
\subsection{Data Pre-processing:}
\indent The data pre-processing stage involves performing data normalization using the Z-score method and minority class oversampling using the SMOTE algorithm. 
\subsubsection{Z-Score Normalization}\mbox{}\\
\indent The first step of the data pre-processing stage is performing Z-score data normalization. However, the data must first be encoded using a label encoder to transform any categorical features into numerical ones. Then, data normalization is performed by calculating the normalized value $x_{norm}$ of each data sample $x_i$ as follows:
\begin{equation}
x_{norm}= \frac{x_i-\mu}{\sigma}
\end{equation}
where $\mu$ being the mean vector of the features and $\sigma$ being the standard deviation. It is worth mentioning that the Z-score data normalization is performed given that ML classification models tend to perform better with normalized datasets \cite{zscore}.
\subsubsection{SMOTE Technique}\mbox{}\\
\indent The second step is performing minority class oversampling using the SMOTE algorithm. This algorithm aims at synthetically creating more instances of the minority class to reduce the class-imbalance that often negatively impacts the ML classification model's performance \cite{traffic_imbalance}. Thus, performing minority class oversampling is important, especially for network traffic datasets which typically suffer from this issue, to improve the training model performance \cite{SMOTE}.\\
\indent Upon analyzing the original minority class instances, SMOTE algorithm synthesizes new instances using the $k$-nearest neighbors concept. Accordingly, the algorithm groups all the instances of the minority class into one set $X_{minority}$. For each instance $X_{inst}$ within $X_{minority}$, a new synthetic instance $X_{new}$ is determined as follows \cite{SMOTE2}:
\begin{equation}
X_{new} = X_{inst} + rand(0,1) * \left(X_{j} - X_{inst}\right), j=1,2,...,k
\end{equation} 
where $rand(0,1)$ is a random value in the range [0,1] and $X_j$ is a randomly selected sample from the set $\left\{X_1,X_2,...,X_k\right\}$ of $k$ nearest neighbors of $X_{inst}$. Note that unlike other oversampling algorithms that replicate minority class instances, SMOTE algorithm generates new high quality instances that statistically resemble samples of the minority class \cite{SMOTE,SMOTE2}.  
\subsection{Feature Selection:}
\indent This work compares between two different feature selection techniques, namely information gain-based and correlation-based feature selection, and explores their effect on the models' detection performance and time complexity. %More specifically, feature selection aims at reducing the complexity of the classification model and consequently decrease the model's training time without sacrificing its performance \cite{FS_reason}. 
This is particularly relevant when designing ML models for large scale systems that generate high dimensional data \cite{FS_reason}. 
\subsubsection{Information Gain-based Feature Selection}\mbox{}\\
\indent The first algorithm considered is the information gain-based feature selection (IGBFS) algorithm. %This algorithm  belongs to the group of ``Information Theory'' feature selection techniques \cite{FS_IG1}. 
As the name suggests, it uses information theory concepts such as entropy and mutual information to select the relevant features \cite{FS_IG1,FS_IG2}. The IGBFS ranks features based on the amount of information (in bits) that can be gained about the target class and selects the ones with the highest amount of information as part of the feature subset provided for the ML model. Thus, the feature evaluation function is \cite{FS_IG2}: 
\begin{equation}
\begin{split}
I(S;C) & = H(S)-H(S|C) \\
& = \sum\limits_{s_i \in S}\sum\limits_{c_j \in C} P(s_i,c_j)log\frac{P(s_i,c_j)}{P(s_i)\times P(c_j)}
\end{split}
\end{equation}
where $I(S;C)$ is the mutual information between feature subset $S$ and class $C$, $H(S)$ is the entropy/uncertainty of discrete feature subset $S$, $H(S|C)$ is the conditional entropy/uncertainty of discrete feature subset $S$ given class $C$, $P(s_i,c_j)$ is the joint probability of feature having a value $s_i$ and class being $c_j$, $P(s_i)$ is the probability of feature having a value $s_i$, and $P(c_j)$ is the probability of class being $c_j$. %The information gained from each feature about the target class is calculated using these values. Then, the ones with the highest amount of information are chosen as part of the feature subset provided for the ML classification model.
\subsubsection{Correlation-based Feature Selection}\mbox{}\\
\indent The second feature selection algorithm considered is the correlation-based feature selection (CBFS) algorithm. %This algorithm belongs to the group of ``Traditional Statistical'' feature selection techniques \cite{FS_CFS1}. 
It is often used due to its simplicity since it ranks features based on their correlation with the target class and selects the highest ones \cite{FS_CFS1,FS_CFS2,sa}. CBFS includes a feature as part of the subset if it is considered to be relevant (\textit{i.e.} if it is highly correlated with or predictive of the class \cite{FS_CFS2,FS_CFS3}). When using CBFS, the Pearson's correlation coefficient is used as the feature subset evaluation function. Thus, the evaluation function is \cite{FS_CFS2}:
\begin{equation}
Merit_S=\frac{k \times \overline{r_{cf}}}{\sqrt{k + k\times(k-1)\times\overline{r_{ff}}}}
\end{equation}
where $Merit_S$ is the merit of the feature subset $S$, $k$ is the number of features in feature subset $S$, $\overline{r_{cf}}$ is the average class-feature Pearson correlation, and $\overline{r_{ff}}$ is the average feature-feature Pearson correlation. %This equation is used to rank the feature subsets with the subset having the highest correlation with the target class being chosen for the ML training model.
\subsection{Hyper-parameter Optimization:}
%\indent As mentioned earlier, the third stage of the framework focuses on optimizing the hyper-parameters of the different ML classification models considered. This is done in an attempt to improve the performance of ML algorithms. This is because each classification algorithm is governed by a set of parameters that dictate its predictive performance \cite{ee}. Therefore, optimizing the ML classification model's hyper-parameters can improve their intrusion detection performance.\\
\indent This work explores different hyper-parameter optimization methods, namely random search (RS), Particle Swarm Optimization (PSO) and Genetic Algorithm (GA) meta-heuristic algorithms, and Bayesian optimization algorithm \cite{sc,ee,hyper1}. %These methods are briefly described in the following subsections.
\subsubsection{Random Search}\mbox{}\\
\indent The first hyper-parameter optimization technique is the RS method. This method belongs to the class of heuristic optimization models \cite{RS_heuristic}. Similar to the grid search algorithm \cite{injadat_ch4,injadat_ch5}, RS tries different combinations of the parameters to be optimized. Mathematically, this translates to the following model:
\begin{equation}
\max\limits_{parm} f(parm)
\end{equation}
where $f$ is an objective function to be maximized (typically the accuracy of the model) and $parm$ is the set of parameters to be tuned. In contrast to the grid search method, the RS method does not perform an exhaustive search of all possible combinations, but rather only randomly chooses a subset of combinations to test \cite{RS_heuristic}. Therefore, RS tends to outperform grid search method, especially when the number of hyper-parameters is small \cite{RS_heuristic}. Additionally, this method also allows for the optimization to be performed in parallel, further reducing its computational complexity \cite{ee}.
\subsubsection{Meta-heuristic Optimization Algorithms}\mbox{}\\
\indent The second class of hyper-parameter optimization methods is the meta-heuristic optimization algorithms. These algorithms aim at identifying or generating a heuristic that may provide a sufficiently good solution to the optimization problem at hand \cite{metaheuristic}. They tend to find suitable solutions for combinatorial optimization problems with a lower computational complexity \cite{metaheuristic}, making them good candidates for hyper-parameter optimization.\\
\indent This work considers two well-known meta-heuristics for hyper-parameter optimization, namely PSO and GA.
\begin{enumerate}
	\item PSO: is a well-known meta-heuristic algorithm that aims at simulating the social behavior such as flocks of birds traveling to a ``promising position'' \cite{PSO1}. In the case of hyper-parameter optimization, the desired ``position'' is the suitable values for the hyper-parameters. In general, PSO algorithm uses a population or a set of particles to search for a suitable solution by iteratively updating these particles' position within the search space.\\ 
	More specifically, each particle looks at its own best previous experience $pbest$ (the cognition part) and the best experience of other particles $gbest$ (the social part) to determine its searching direction change. Mathematically, the position of the particle at each iteration $t$ is represented as a vector $x_i^t=\{x_{i1}^t, x_{i2}^t,...,x_{iD}^t\}$ and its velocity as $v_i^t=\{v_{i1}^t, v_{i2}^t,...,v_{iD}^t\}$ where $D$ is the number of parameters to be optimized. Assuming that $pbest_i^t$ is particle $i$'s best solution until iteration $t$ and $gbest^t$ is the best solution within the population at iteration $t$, each particle changes its velocity as follows \cite{PSO1}:
	\begin{equation}
	%\begin{split}
	%&
	v_{id}^t=v_{id}^{t-1}+c_1r_1(pbest_{id}^t-x_{id}^t)+c_2r_2(gbest_{d}^t-x_{id}^t) \\ 
	%& d=1,2,...D
	%\end{split}
	\end{equation}
	where $c_1$ is the particle's cognition learning factor, $c_2$ the social learning factor, and $r_1$ and $r_2$ being random numbers between [0,1]. Accordingly, the particle's new position becomes \cite{PSO1}:
	\begin{equation}
	x_{id}^{t+1}=x_{id}^t + v_{id}^t %\; d=1,2,...D
	\end{equation}
	Within the context of hyper-parameter optimization, $x_i^t=parm$ where $parm$ is the set of parameters for the ML model under consideration. For example, in the case of SVM, the parameters are $C$ and $\gamma$.      
	\item GA: is another well-known meta-heuristic algorithm that is inspired by the evolution and the process of natural selection \cite{GA1}. It is often used to identify high-quality solutions to combinatorial optimization problems using biologically inspired operations including mutation, crossover, and selection \cite{GA1}. Using these operators, GA algorithms can search the solution space efficiently \cite{GA1}.\\
	In the context of ML hyper-parameter optimization, GA algorithm works as follows \cite{GA1}:
	\begin{enumerate}
		\item[a)] Initialize a population of random solutions denoted as chromosomes. Each chromosome is a vector of potential hyper-parameter value combinations.
		\item[b)] Determine the fitness of each chromosome using a fitness function. The function is typically the ML model's accuracy when using each chromosome's vector.
		\item[c)] Rank the chromosomes according to their relative fitness in descending order.
		\item[d)] Replace least-fit chromosomes with new chromosomes generated through crossover and mutation processes. %Crossover refers to the process of generating new off-springs from two parent chromosomes by exchanging their ``genes'' (each gene represents a value for one of the hyper-parameters). On the other hand, mutation refers to the processing of altering a chromosome randomly. Within this process, a gene is randomly selected and altered.
		\item[e)] Repeat steps b)-d) until the performance is no longer improving or some stopping criterion is met. 
	\end{enumerate}
	%It is worth mentioning that each chromosome is represented as an encoded string of bits of length $l$ which depends on the range of values for each hyper-parameter.\\
	Due to its effectiveness in identifying very good solutions (near-optimal in many cases), this meta-heuristic has been used in a variety of applications including workflow scheduling \cite{GA_app1}, photovoltaic systems \cite{GA_app2}, wireless networking \cite{GA_app3}, and in this case machine learning \cite{GA_app4}. 
\end{enumerate}
\subsubsection{Bayesian Optimization}\mbox{}\\
\indent The third hyper-parameter optimization method considered in this work is the Bayesian Optimization method. This method belongs to the class of probabilistic global optimization models \cite{BO1}. This method aims at minimizing a scalar objective function $f(x)$ for some value $x$. The output of this optimization process for the same input $x$ differs based on whether the function is deterministic or stochastic \cite{bib23}. The minimization process is divided into  three main parts: a surrogate model that fits all the points of the objective function $f(x)$, a Bayesian update process that modifies the surrogate model after each new evaluation of the objective function, and an acquisition function $a(x)$. Different surrogate models can be assumed, namely the Gaussian Process and the Tree Parzen Estimator.
\begin{enumerate}
	\item Gaussian Process (GP): The model is assumed to follow a Gaussian distribution. Thus, it is of the form \cite{BO2}:
	\begin{equation}
	p(f(x)\big|\;x,parm)= N(f(x)\big|\;\hat{\mu},\hat{\sigma}^2)
	\end{equation}
	where $parm$ is the configuration space of the hyper-parameters and $f(x)$ the value of the objective function with $\hat{\mu}$ and $\hat{\sigma}^2$ being its mean and variance respectively. Note that such a model is effective when the number of hyper-parameters is small, but is ineffective for conditional hyper-parameters \cite{BO3}.
	\item Tree Parzen Estimator (TPE): The model is assumed to follow one of two density functions, $l(x)$ or $g(x)$ depending on some pre-defined threshold $f^{*}(x)$ \cite{BO2}:
	\begin{equation}
	p(x\big|\;f(x),parm) = \begin{cases}
	l(x) &\text{if $f(x) < f^{*}(x)$}\\
	g(x) &\text{if $f(x) > f^{*}(x)$}
	\end{cases}
	\end{equation}
	where $parm$ is the configuration space of the hyper-parameters and $f(x)$ the value of the objective function. Note that TPE estimators follow a tree-structure and can optimize all hyper-parameter types \cite{BO3}. 
\end{enumerate}
Based on the surrogate model assumption, the acquisition function is maximized to determine the subsequent evaluation point. The role of the function is to measure the expected improvement in the objective while avoiding values that would increase it \cite{bib23}. Therefore, the expected improvement (EI) can be determined as follows:
\begin{equation}
EI(x,Q)=E_{Q}\big[\max(0,\mu_{Q}(x_{best})-f(x))\big]
\end{equation}
where $x_{best}$ is the location of the lowest posterior mean and $\mu_{Q}(x_{best})$ is the lowest value of the posterior mean.
\begin{figure}[!t]
	\centering
	\includegraphics[trim=2cm 0cm 1cm 0cm,scale=.5]{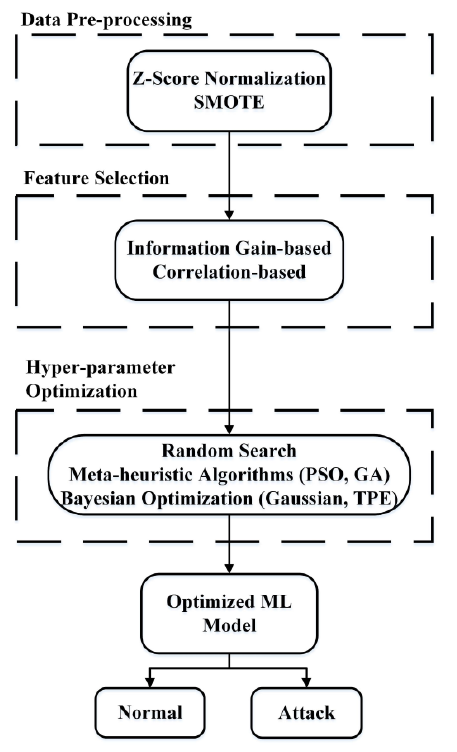}
	\caption{Proposed Multi-stage Optimized ML-based NIDS Framework}
	\label{approach_fig}
\end{figure}
\section{Proposed Multi-Stage Optimized ML-based NIDS Framework}\label{proposed_framework}
\subsection{General Framework Description:}
\indent This work focuses on building a multi-stage optimized ML-based NIDS framework that achieves high detection accuracy, low FAR, and has a low time complexity.The proposed framework is divided into three main stages to achieve this goal. The first stage includes the data pre-processing that includes performing Z-score normalization and Synthetic Minority Oversampling TEchnique (SMOTE). This is done to improve the performance of the training model and reduce the class-imbalance often observed in network traffic data \cite{traffic_imbalance}. In turn, this can reduce the training sample size since the ML model would have enough samples to understand the behavior of each class \cite{SMOTE}.\\ 
\indent The second stage of the proposed framework is conducting a feature selection process to reduce the number of features needed for the ML classification model. This is done to reduce the time complexity of the classification model and consequently decrease its training time without sacrificing its performance \cite{FS_reason}. With that in mind, two different methods are compared within this stage of the framework.\\ 
\indent The third stage of the framework involves the optimization of the hyper-parameters of the different ML classification models considered. To that end, three different hyper-parameter tuning/optimization models are investigated, namely random search, meta-heuristic optimization algorithms including particle swarm optimization (PSO) and genetic algorithm (GA), and Bayesian Optimization (BO) algorithm. These models represent three different hyper-parameter tuning/optimization categories which are heuristics \cite{RS_heuristic}, meta-heuristics \cite{evolutionary1}, and probabilistic global optimization \cite{BO1} models respectively.\\
\indent The results of these optimization stages are combined to build the optimized ML classification model for effective NIDS system that classifies new instances as either normal or attack instances. Figure \ref{approach_fig} illustrates the different stages of the proposed framework.
\subsection{Security Considerations:}
\indent The proposed multi-stage optimized  ML-based NIDS framework is a signature-based NIDS system. This is illustrated by the fact that the framework oversamples the minority class, which typically is the attack class in network traffic \cite{sf,sf1}. Thus, the framework learns from the observed patterns of the known initiated attacks \cite{sf,sf1}. However, it is worth noting that the framework can work as an anomaly-based NIDS since it is trained by adopting a binary classification model so that it can classify any anomalous behavior as an attack.\\ 
\indent This framework can be deployed as one module within a more comprehensive security framework/policy that an individual or organization can adopt. This security framework/policy can include other mechanisms such as firewalls, deep packet inspection, user access control, and user authentication mechanisms \cite{sdp1}\cite{sdp2}. This would offer a multi-layer secure framework that can preserve the privacy and security of the users' data and information. 
\subsection{Complexity:}
\indent To determine the time complexity of the proposed multi-stage optimized ML-based NIDS framework, we need to determine the complexity of each algorithm used in each stage. Given that this work compares the performance of different algorithms within the different stages of the framework, the overall time complexity is determined by the combination of algorithms that results in the highest aggregate complexity.\\
\indent It is assumed that the data is composed of $M$ samples and $N$ features. Starting with the first stage, \textit{i.e.} the data pre-processing stage, the complexity of the Z-score normalization process is $O(N)$ since we need to normalize all the samples of the $N$ features within the dataset. On the other hand, the complexity of the SMOTE algorithm is $O(M^2_{min} N)$ where $M_{min}$ is the number of samples belonging to the minority class \cite{SMOTE_complexity}. Thus, the overall complexity of the first stage is  $O(M^2_{min} N)$. \\ 
\indent The complexity of the second stage is dependent on the complexity of the different feature selection algorithms considered. The complexity of Correlation-based feature selection is $O(MN^2)$ since this method needs to calculate all the class-feature and feature-feature correlations \cite{FS_CFS2}. In contrast, the complexity of the information gain-based feature selection method is $O(MN)$. This is due to the fact that this method has to calculate the joint probabilities of the class-feature interaction \cite{FS_IG2}. Therefore, the overall complexity of the second stage is $O(MN^2)$.\\
\indent Similarly, the complexity of the third stage depends on the complexity of each of the hyper-parameter optimization methods and the underlying ML model. Starting with the RS method, its complexity is $O(N_{parm}logN_{parm})$ where $N_{parm}$ is the number of parameters to be optimized \cite{RS_complexity}. Conversely, the complexity of the PSO algorithm is $O(N_{parm} N_{pop})$ where $N_{pop}$ is the population size, \textit{i.e.} the number of swarm particles or potential solutions that we start with \cite{PSO_complexity}. In a similar fashion, it can be shown that the complexity of the GA algorithm is also $O(N_{parm} N_{pop})$ where $N_{pop}$ is the population size, \textit{i.e.} the number of chromosomes/potential solutions at the initialization stage \cite{GA_complexity}. For the GP-based BO algorithm, the complexity is $O(M^3_{red})$ where $M_{red}$ is the size of the reduced training sample. This is because the optimization process is carried on the training sample chosen after pre-processing and feature selection. In contrast, the time complexity of the TPE-based BO model is $O(M_{red}logM_{red})$ since this model follows a tree-like structure when performing the optimization \cite{BO_TPE_complexity}. \\
\indent Based on the aforementioned discussion, the overall complexity of the proposed framework is $O(MN^2)$. This is because the second stage will dominate the complexity as it would still use the complete dataset rather than the reduced training dataset. As such, even if we consider the complexity of the potential ML classification model (for example the complexity of KNN classifier can be estimated as $O(M_{red}N_{red})$ \cite{complexity1,complexity2} where $N_{red}$ is the size of the reduced feature set), it is dependent on the reduced training sample dataset with reduced feature set size. Hence, the multi-stage optimized ML-based NIDS framework's complexity is $O(MN^2)$. Determining the overall time complexity of the complete framework including the optimized ML model training is essential since the model will be frequently re-trained to learn new attack patterns. This is based on the fact that network intrusion attacks continue to evolve and thus organizations need to have a flexible and dynamic NIDSs to keep up with these new attacks.  
\section{Datasets Description}\label{datasets}
\indent This work uses two state-of-the-art intrusion datasets to evaluate the performance of the proposed multi-stage optimized ML-based NIDS framework. In what follows, a brief description of the two datasets is given.
\begin{figure}[!b]
	\centering
	\includegraphics[trim=2cm 0cm 1cm 1cm,scale=0.45]{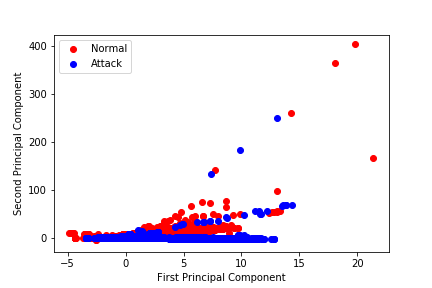}
	\caption{Principal Component Analysis of CICIDS 2017 Dataset Illustrating its Non-linear Nature}
	\label{cicids_pca}
\end{figure}
\subsection{CICIDS 2017}
\indent The first dataset under consideration is the Canadian Institute of Cybersecurity's IDS 2017 (CICIDS2017) dataset \cite{cicids17}. This dataset is an extension of the ISCX 2012 dataset used in our previous work \cite{sc}. The dataset was generated with the goal of it resembling realistic background traffic \cite{cicids17}. As such, the dataset contains benign and 14 of the most up-to-date common network attacks. The data collection process span a duration of five days from Monday July 3 till Friday July 7, 2017. Within this period, different attacks where generated during different time windows. The resulting dataset contained \textbf{3,119,345 instances} and \textbf{83 features} (1 class feature and 82 statistical features) representing the different characteristics of a network traffic request such as duration, protocol used, packet size, as well as source and destination details. However, nearly 300,000 samples were unlabeled and hence were discarded. Therefore, the refined dataset considered in this work contains \textbf{2,830,540 instances in total} with \textbf{2,359,087} being \textbf{BENIGN} and \textbf{471,453} being \textbf{ATTACK}. Note that the attack instances represent various types of real-world network traffic attacks such as denial-of-service (DoS) and port scanning. However, this work merged all attacks into one label as the goal is to detect an attack regardless of its nature.\\
\indent Fig. \ref{cicids_pca} shows the first and second principal components for the CICIDS 2017 dataset. It can be seen that the two classes are intertwined. Moreover, it can be observed that the features of the dataset are non-linear. Hence, we would expect a non-linear kernel to perform better in classifying the instances of this dataset.
\begin{figure}[!b]
	\centering
	\includegraphics[trim=2cm 0cm 1cm 1cm,scale=0.45]{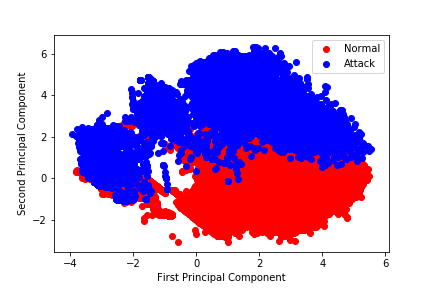}
	\caption{Principal Component Analysis of UNSW-NB 2015 Dataset Illustrating its Non-linear Nature}
	\label{unsw_pca}
\end{figure}   
\subsection{UNSW-NB 2015}
\indent The second dataset considered is the University of New South Wales's network intrusion dataset (UNSW-NB 2015) generated in 2015 \cite{unsw15}. The dataset is a hybrid of real modern network normal activities and synthetic attack behaviors \cite{unsw15}. The data was collected through two different simulations conducted on two different days, namely January 22 and February 17, 2015. The resulting dataset consists of \textbf{2,540,044 instances} and \textbf{49 features} (1 class feature and 48 statistical features) representing the different characteristics of a network traffic request such as source and destination details, duration, protocol used, and packet size \cite{unsw15}. These instances are labeled as follows: \textbf{2,218,761 normal instances} and \textbf{521,283 attack instances}. In this case, no merging of attacks was needed since the dataset was originally labeled in a binary fashion.\\
\indent In a similar fashion, Fig. \ref{unsw_pca} shows the first and second principal components for the UNSW-NB 2015 dataset. Again, we can observe that the features are non-linear. However, it can be observed that the level of intertwining between the two classes is lower. Accordingly, it is easier to separate between the two classes.\\
\indent Note that there are other network intrusion detection datasets that can be studied such as the NSL KDD 99 dataset and the Kyoto 2006+ datasets. However, these datasets have already been extensively studied. Moreover, they are outdated and may not have recent attack patterns. In contrast, the two datasets considered in this work are more recent and have more attack patterns. As such, studying them will provide better equipped NIDSs that are trained to detect more attack types.
\subsection{Attack Types}
\indent The two datasets considered in this work contain some similar attacks and some that are different. For example, the CICIDS 2017 dataset contains the following attacks: Denial-of-Service (DoS), port scanning, brute-force, web-attacks, botnets, and infiltration \cite{cicids17}. In contrast, the UNSW-NB 2015 dataset contains the following attacks: fuzzers, analysis, backdoors, DoS, exploits, generic, reconnaissance, shellcode, and worms \cite{unsw15}. Accordingly, it can be deduced that the proposed framework learns the patterns of various attack types.\\
\indent Note that the proposed framework adopts a binary classification model by labeling all attack types as ``attack''. The goal is to develop a NIDS that can detect various attacks rather than just a finite group of common attacks such as DoS. This reiterates the idea that the proposed multi-stage optimized ML-based NIDS can work as an anomaly-based NIDS despite its training as a signature-based NIDS. 
\section{Experimental Performance Evaluation}\label{performance}
\subsection{Experimental Setup}\label{exp_setup}
\indent The experiments conducted for this work were completed using \textbf{Python 3.7.4} running on Anaconda's Jupyter Notebook. This was run on a virtual machine having a 3 processors Intel (R) Xeon (R) CPU E5-2660 v3 \@ 2.6 GHz and 64GB of memory running Windows Server 2016. The experimental results are divided into three main subsections, namely the impact of data pre-processing on training sample size, impact of feature selection on feature set size and training sample size, and the impact of optimization methods on the ML models' detection performance.\\
\indent The classification models used in this work are KNN classifier and the RF classifier. These classifiers were chosen due to two main reasons. Firstly, these classifiers were the top performing classifiers in our previous work as they showed their effectiveness with network intrusion detection \cite{sc}. Secondly, these classifiers have lower computational complexities when compared to other classifiers. For example, the KNN classifier has a complexity of $O(MN)$ where $M$ is the number of instances and $N$ is the number of features \cite{complexity1,complexity2}. Similarly, the complexity of the RF classifier is $O(M^2\sqrt{N}t)$ where $t$ is the number of trees within the RF classifier. However, since this classifier allows for multi-threading, its training time is significantly reduced to approximately $O(\frac{N^2\sqrt{M}t}{threads})$ where $threads$ is the maximum number of participating threads \cite{sd}. In contrast, the complexity of SVM can reach an order of $O(M^3N)$ \cite{svm_complexity}. Therefore, training such a model would be computationally prohibitive, especially given the dataset sizes used in this work. Note that the parameters to be tuned are: 
\begin{itemize}
	\item KNN: number of neighbors K.
	\item RF: Splitting criterion (Gini or Entropy) and Number of trees.
\end{itemize}
\indent It is worth noting that the runtime complexity (also commonly referred to as testing complexity) of KNN and RF optimized models is $O(MN)$ and $O(Nt)$ respectively where $M$ is the number of training samples, $N$ is the number of features, and $t$ is the number of decision trees forming the RF classifier \cite{KNN_testing_complexity,RF_testing_complexity}. In the case of KNN, any new instance is classified after calculating the distance between itself and all other instances in the training sample and identifying its K nearest neighbors \cite{KNN_testing_complexity}. Conversely, when using the RF classifier, the new instance is fed to the $t$ different decision trees, each of which uses $N$ splits based on the $N$ features considered, and the class is determined based on the majority vote among these $t$ trees. 
\subsection{Results and Discussion}\label{results}
\subsubsection{Impact of data pre-processing on training sample size}
\begin{figure}[!t]
	\centering
	\includegraphics[trim=1cm 0cm 1cm 1cm,scale=.28]{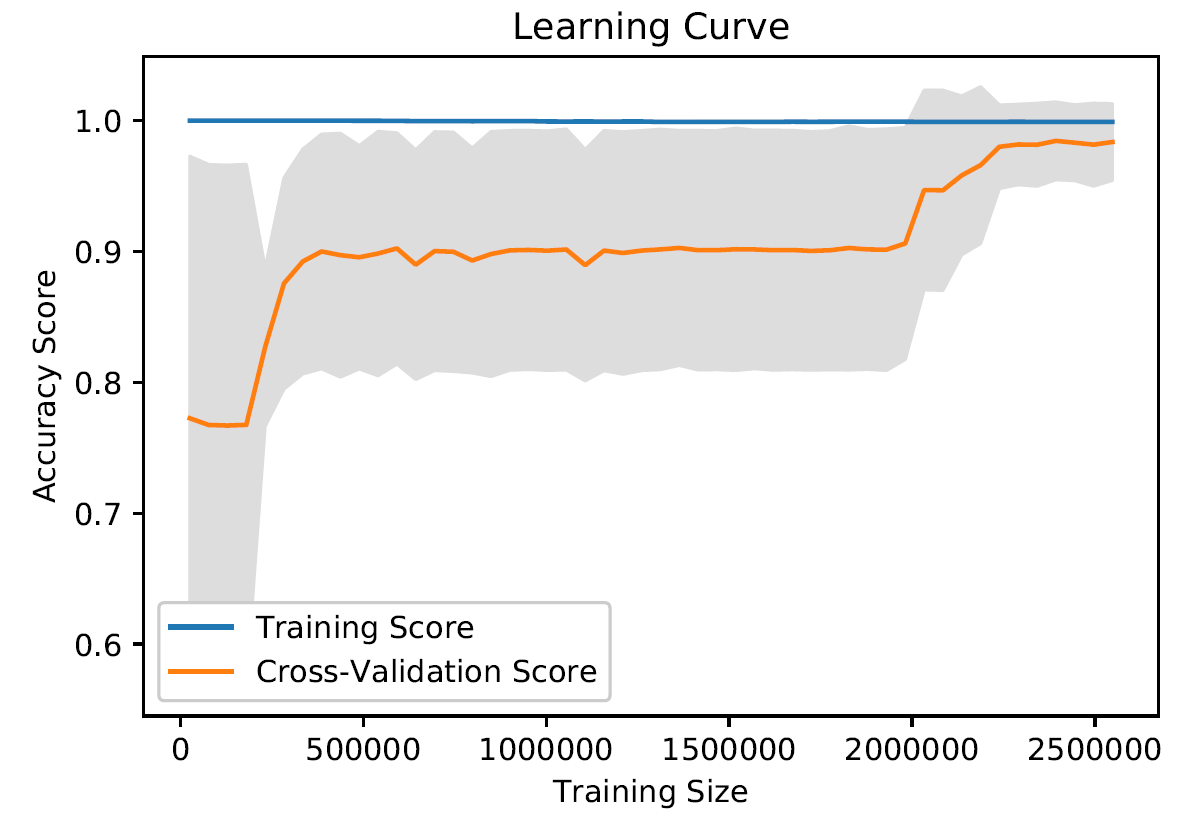}
	\caption{Learning Curve Showing Training and Cross-Validation Accuracy for CICIDS 2017 Dataset Before SMOTE}
	\label{cicids_no_smote}
\end{figure}

\begin{figure}[!t]
	\centering
	\includegraphics[trim=1cm 0cm 1cm 1cm,scale=.28]{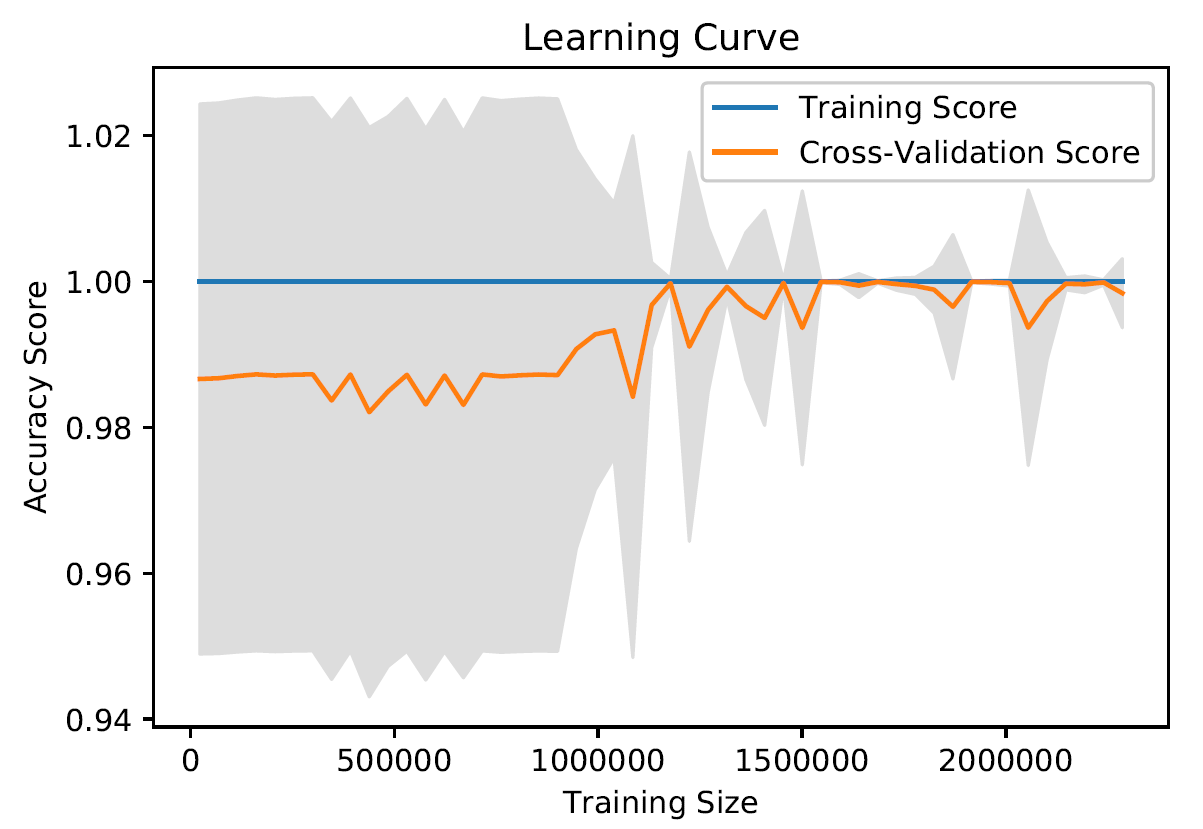}
	\caption{Learning Curve Showing Training and Cross-Validation Accuracy for UNSW-NB 2015 Dataset Before SMOTE}
	\label{unsw_no_smote}
\end{figure}

\mbox{}\\
\indent Starting with the impact of data pre-processing stage on the training sample size, the learning curve showing the variation of training accuracy and the cross-validation accuracy as the training sample size changes. Both datasets were split randomly into training and testing samples after normalization using a 70\%/30\% split criterion.\\
\indent Using the SMOTE technique, the number of instances of each type in each dataset's training sample is as follows:
\begin{itemize}
	\item CICIDS 2017: 1,818,477 \textbf{benign} instances (denoted as 0) and 1,800,000 \textbf{attack} instances (denoted as 1).
	\item UNSW-NB 2015: 1,775,010 \textbf{normal} instances (denoted as 0) and 1,500,000 \textbf{attack} instances (denoted as 1). 
\end{itemize}
It can be seen from Fig. \ref{cicids_no_smote} that the number of training samples needed for the CICIDS 2017 dataset for the training accuracy and cross-validation accuracy to converge is close to 2.3 million samples. Similarly, for the UNSW-NB 2015 dataset, the number of training samples needed is close to 1.3 million samples as can be seen from Fig. \ref{unsw_no_smote}. This can be attributed to the fact that both datasets are originally imbalanced with much fewer attack samples when compared to normal samples. Hence, the model struggles to learn the attack patterns and behaviors. \\    
\begin{figure}[!t]
	\centering
	\includegraphics[trim=1cm 0cm 1cm 1cm,scale=.28]{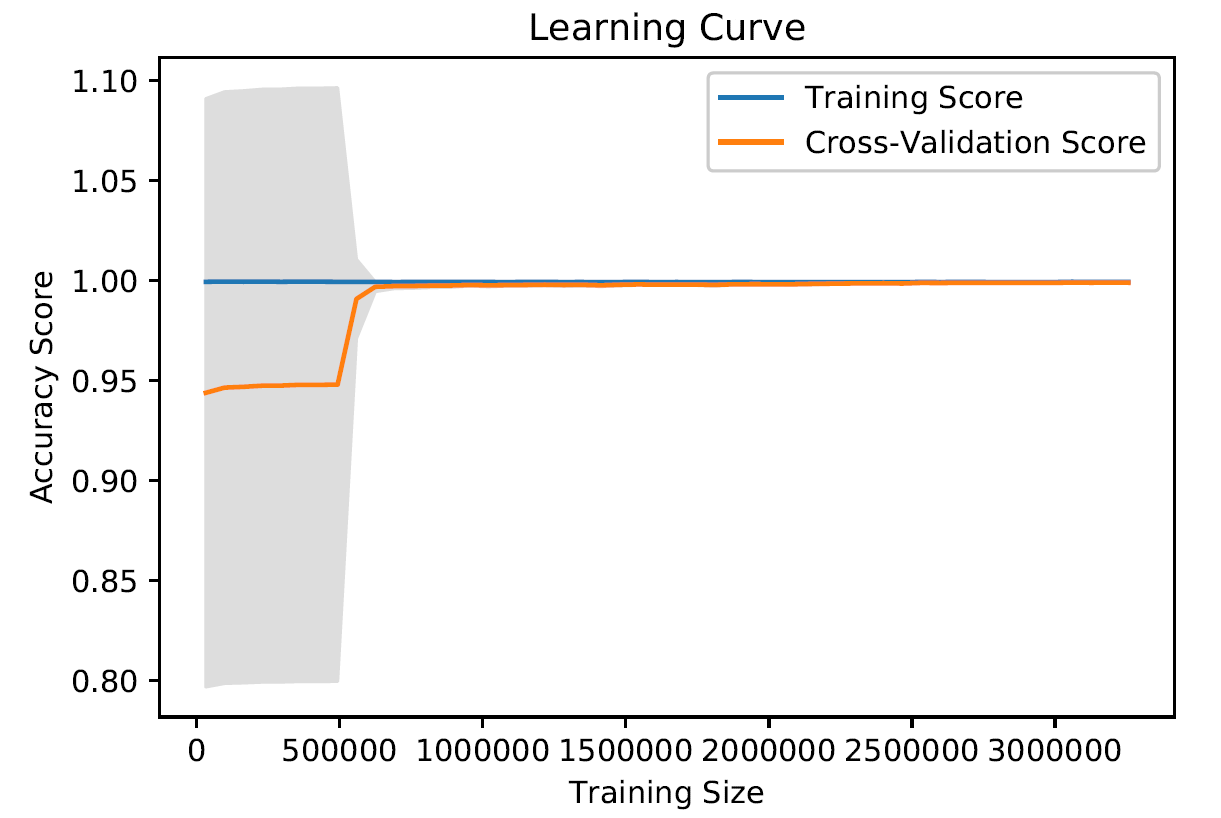}
	\caption{Learning Curve Showing Training and Cross-Validation Accuracy for CICIDS 2017 Dataset After SMOTE}
	\label{cicids_smote}
\end{figure}
\begin{figure}[!t]
	\centering
	\includegraphics[trim=1cm 0cm 1cm 1cm,scale=.28]{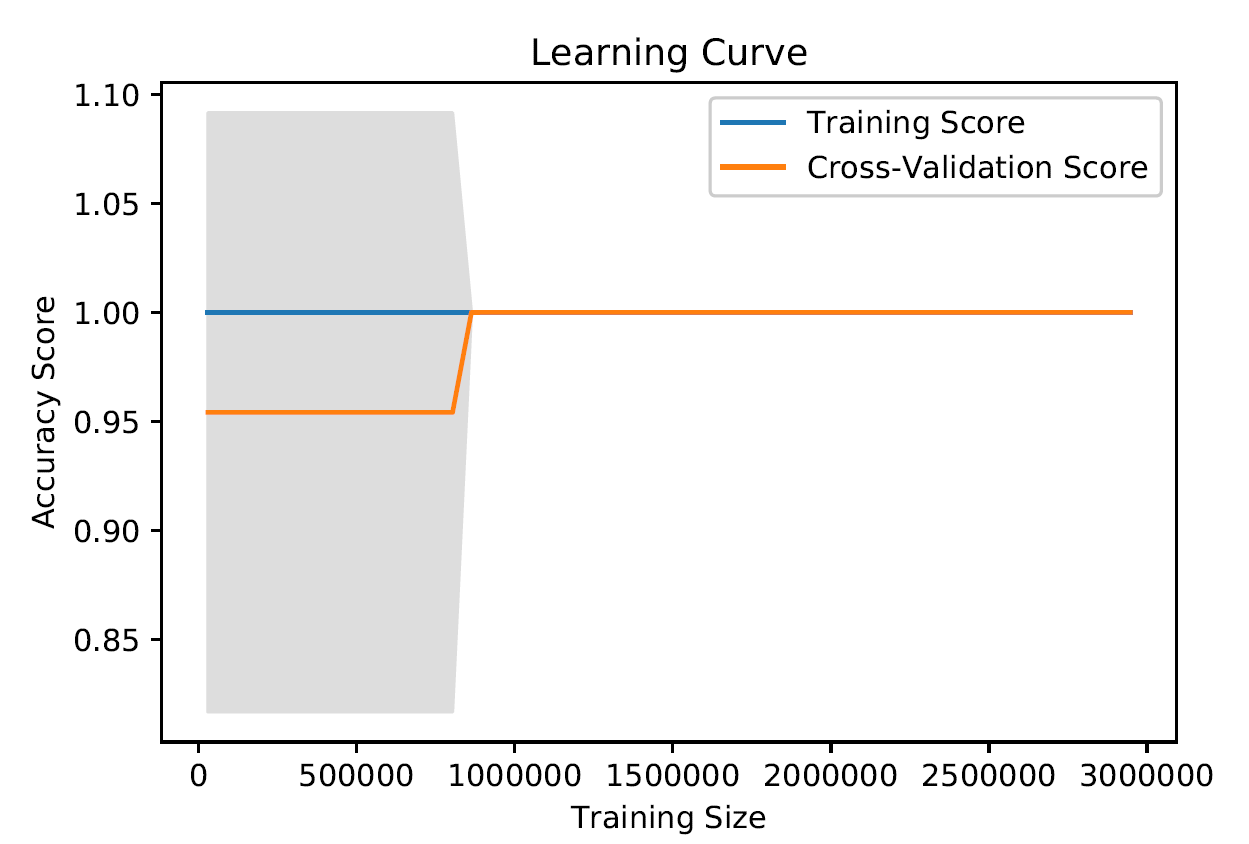}
	\caption{Learning Curve Showing Training and Cross-Validation Accuracy for UNSW-NB 2015 Dataset After SMOTE}
	\label{unsw_smote}
\end{figure}
\indent In contrast, it can be seen from Figs. \ref{cicids_smote} and \ref{unsw_smote} that the number of training samples needed is around 600,000 samples and 800,000 samples for the CICIDS 2017 and UNSW-NB 2015 respectively. This represents a drop of approximately 74\% and 39\% in the training sample size for the two datasets respectively. This highlights the positive impact of using SMOTE technique as it was able to significantly reduce the size of the training sample needed without sacrificing the detection performance. This is mainly due to the introduction of more attack samples that allow the ML model to better learn their patterns and behaviors. To further highlight the impact of using data pre-processing phase, the time needed to build the learning curve was determined. For example, building the learning curve for the UNSW-NB 2015 dataset needed close to 600 minutes prior to applying SMOTE. In contrast, it required around 90 minutes after implementing SMOTE. This highlights the time complexity reduction associated with adopting an oversampling technique.\\
\indent Moreover, it can be seen from all these figures that the models developed before and after SMOTE for both datasets do not suffer from overfitting as illustrated by the relatively small error gap between the training and cross-validation accuracy in Figs. \ref{cicids_no_smote} and \ref{unsw_no_smote} and the zero error gap seen in Figs. \ref{cicids_smote} and \ref{unsw_smote}. As per \cite{overfitting_indicator}, overfitting can be observed from the learning curve whenever the error gap between the training accuracy and the cross-validation accuracy is large. Thus, a small or zero error gap implies that the developed model is not too specific to the training dataset but can perform equally well on the testing and cross-validation sets. 

\begin{figure*}[!t]
	\centering
	\includegraphics[trim=3cm 0cm 1cm 0cm,scale=.45]{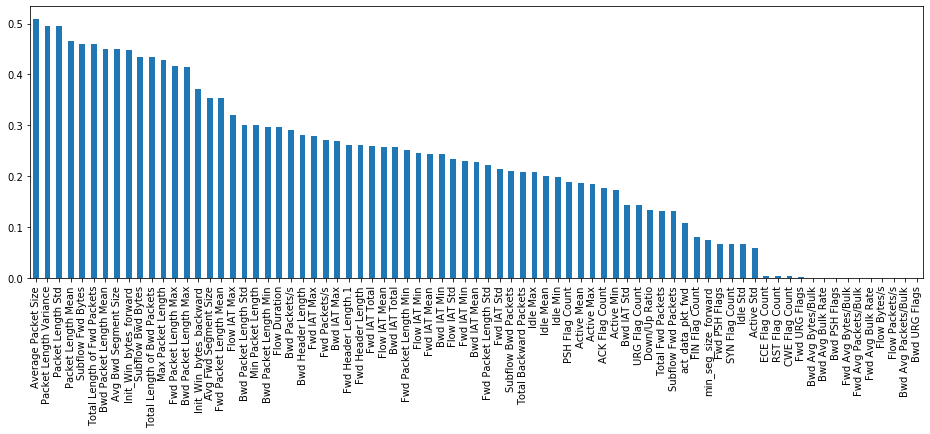}
	\caption{Mutual Information Score of Features for CICIDS 2017 Dataset Showing the Highest Scoring Features in Descending Order}
	\label{cicids_mi_score}
\end{figure*}
\begin{figure*}[!t]
	\centering
	\includegraphics[trim=3cm 0cm 1cm 0cm,scale=.45]{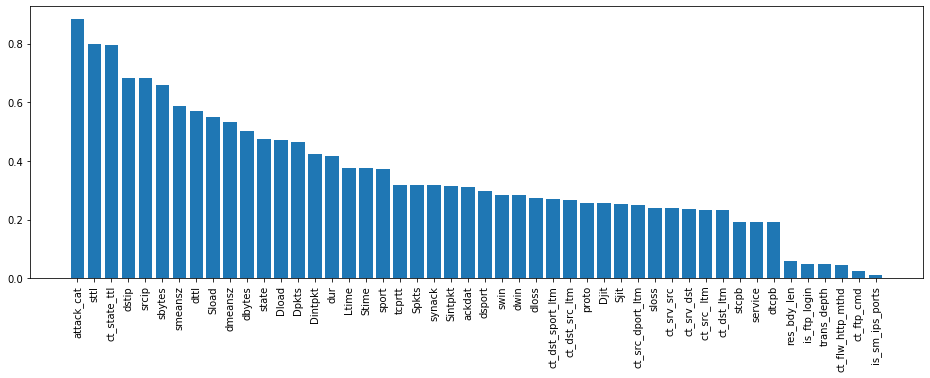}
	\caption{Mutual Information Score of Features for UNSW-NB 2015 Dataset Showing the Highest Scoring Features in Descending Order}
	\label{unsw_mi_score}
\end{figure*}
\subsubsection{Impact of feature selection on feature set size and training sample size}\mbox{}\\
The second stage of analysis involves studying the impact of the different feature selection algorithms on the feature set size and training sample size.
\begin{enumerate}
	\item \textit{Impact of feature selection on feature set size:} Starting with the IGBFS method, Figs. \ref{cicids_mi_score} and \ref{unsw_mi_score} show the mutual information score for each of the features for the CICIDS 2017 and UNSW-NB 2015 datasets respectively. For example, for the CICIDS 2017 dataset, some of the most informative features include the average packet size and packet length variance. Similarly, for the UNSW-NB 2015 dataset, some of the most informative features are also the packet size (denoted by sbyte and dbyte features) and the time to live values. This illustrates the tendency of attacks to have different packet sizes when compared to normal traffic. Moreover, the figures also show that some IPs may have a higher tendency to initiate attacks, which means they are more likely to be compromised.\\ 
	Based on the figures, the number of features  selected for the CICIDS 2017 and UNSW-NB 2015 datasets is 31 features and 19 features, respectively. This represents a reduction of 62\% and 61\% in the feature set size for the two datasets respectively. This is caused by the IGBFS method choosing the relevant features that provide the most information about the class.\\ 
	In contrast, when using the CBFS method, the number of selected features for the CICIDS 2017 and UNSW-NB 2015 datasets is 41 and 32 features respectively. This represents a reduction of 50\% and 33.3\% for each of the datasets, respectively. This reduction is due to the CBFS method choosing the relevant features that are highly correlated with the class feature, \textit{i.e.}  the features whose variation is also reflected in a variation in the corresponding class.\\
	The IGBFS method tends to choose a lower number of features when compared to the CBFS method. This is because the CBFS method relies on the correlation. Thus, two features may be chosen that are highly correlated with the class because they have a high correlation between them and one of them is highly correlated with the class. On the other hand, the IGBFS method studies the features one by one with respect to the class and selects the features that provide the highest amount of information about the class without considering the mutual information between the features themselves. Hence, a lower number of features is typically chosen by the IGBFS method.    
\item \textit{Impact of feature selection on training sample size:} In addition to the impact of the feature selection process on the feature set size, this work also studies its impact on the training sample size. Starting with the IGBFS method, it can be seen from Figs. \ref{cicids_information} and \ref{unsw_information} that the training sample size was reduced to 250,000 and 110,000 samples for the CICIDS 2017 and UNSW-NB 2015 datasets, respectively. This represents a reduction of 59\% and 86\% when compared to the required training sample size after the SMOTE technique is applied. This shows that the IGBFS method can keep the features that provide the most information about the class and discard any feature that may be negatively impacting the learning process.\\ 
\begin{figure}[!t]
	\centering
	\includegraphics[trim=1cm 0cm 1cm 0cm,scale=.28]{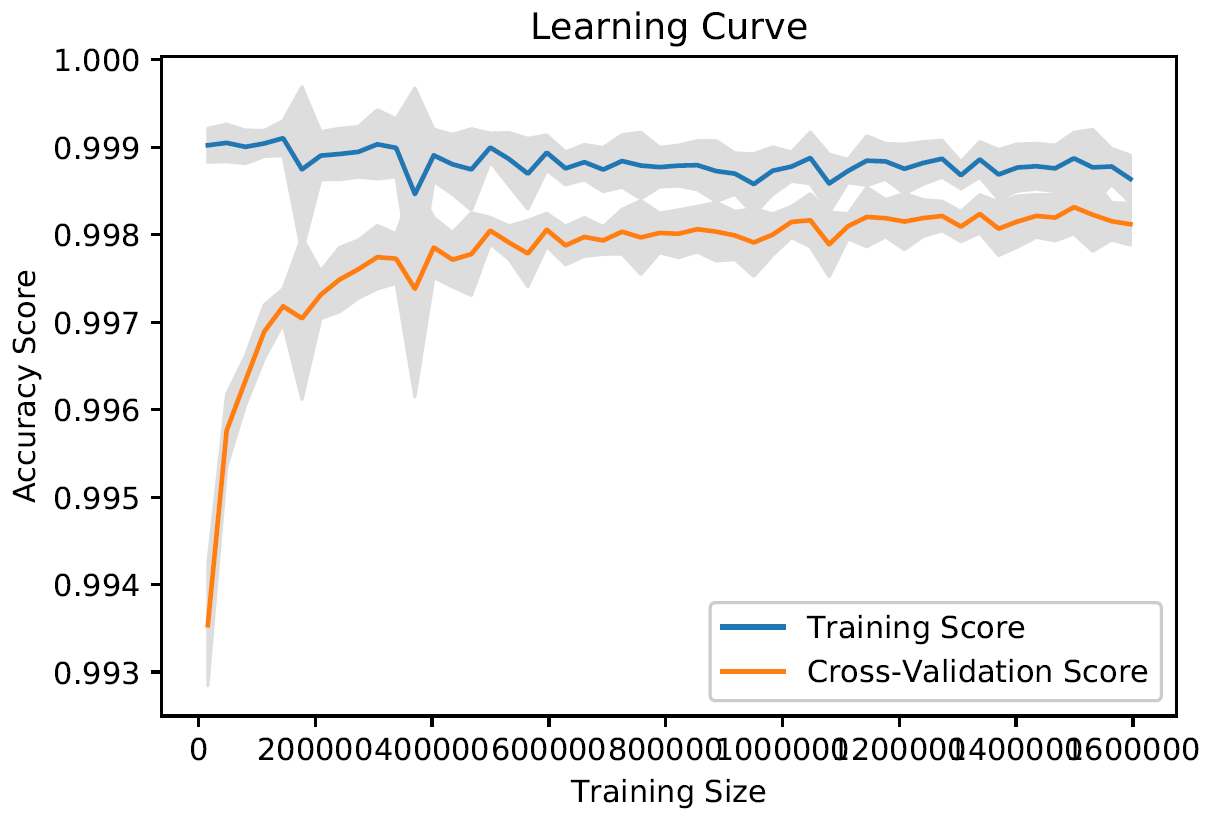}
	\caption{Learning Curve Showing Training and Cross-Validation Accuracy for CICIDS 2017 Dataset After IGBFS}
	\label{cicids_information}
\end{figure}
\begin{figure}[!t]
	\centering
	\includegraphics[trim=0cm 0cm 1cm 0cm,scale=.28]{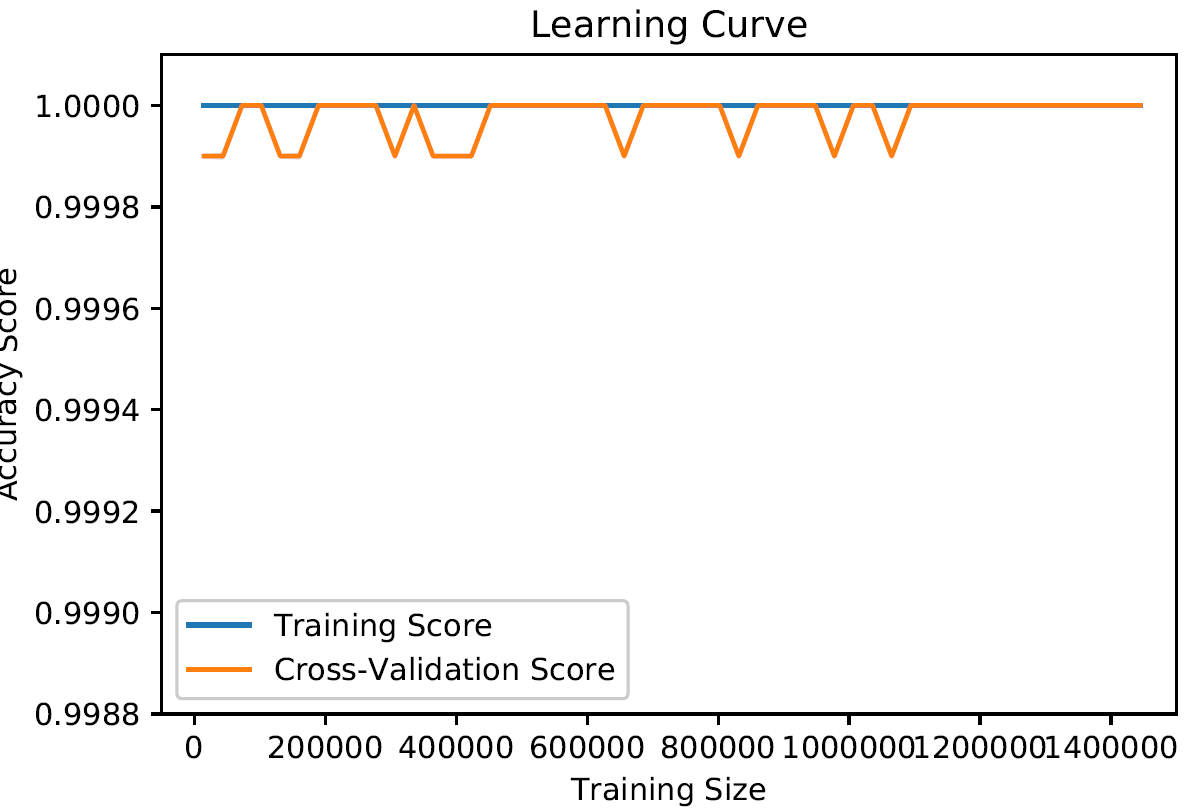}
	\caption{Learning Curve Showing Training and Cross-Validation Accuracy for UNSW-NB 2015 Dataset After IGBFS }
	\label{unsw_information}
\end{figure}
\begin{figure}[!t]
	\centering
	\includegraphics[trim=1cm 0cm 1cm 0cm,scale=.28]{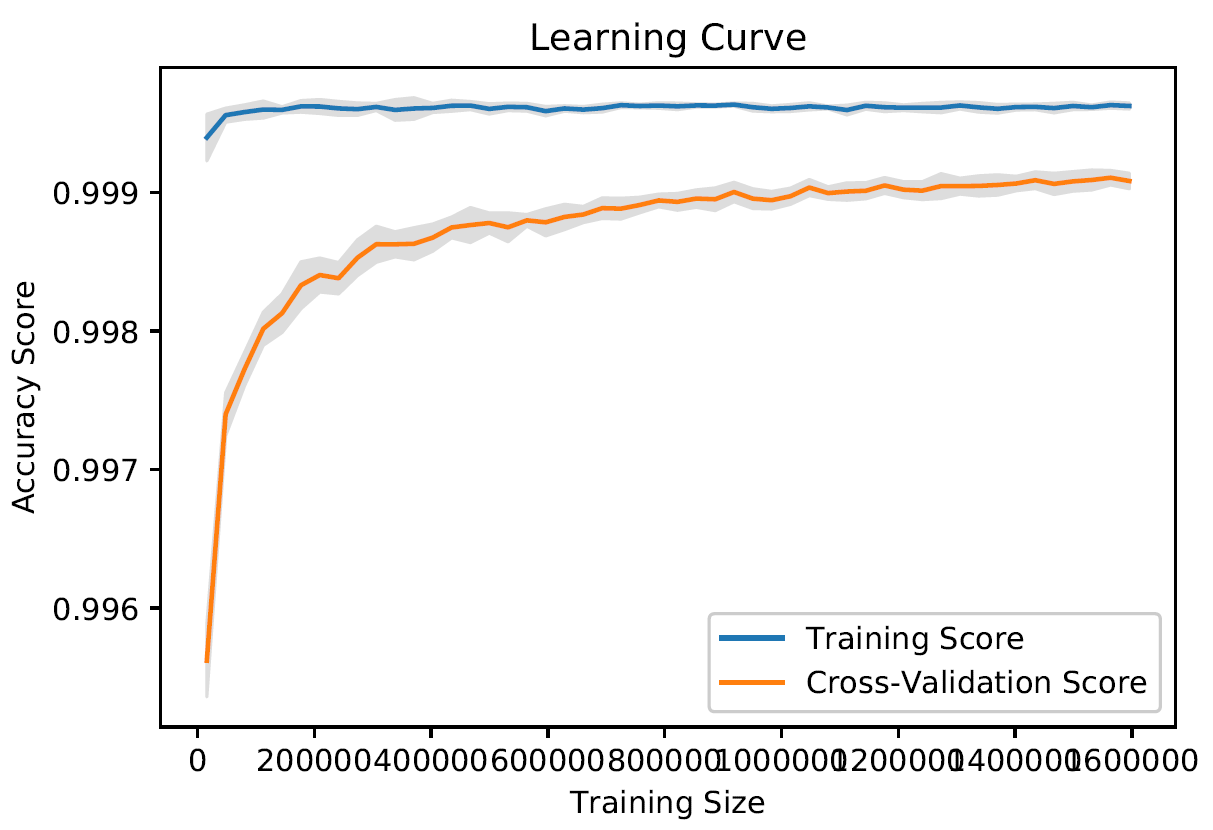}
	\caption{Learning Curve Showing Training and Cross-Validation Accuracy for CICIDS 2017 Dataset After CBFS }
	\label{cicids_correlation}
\end{figure}
\begin{figure}[!t]
	\centering
	\includegraphics[trim=1cm 0cm 1cm 0cm,scale=.28]{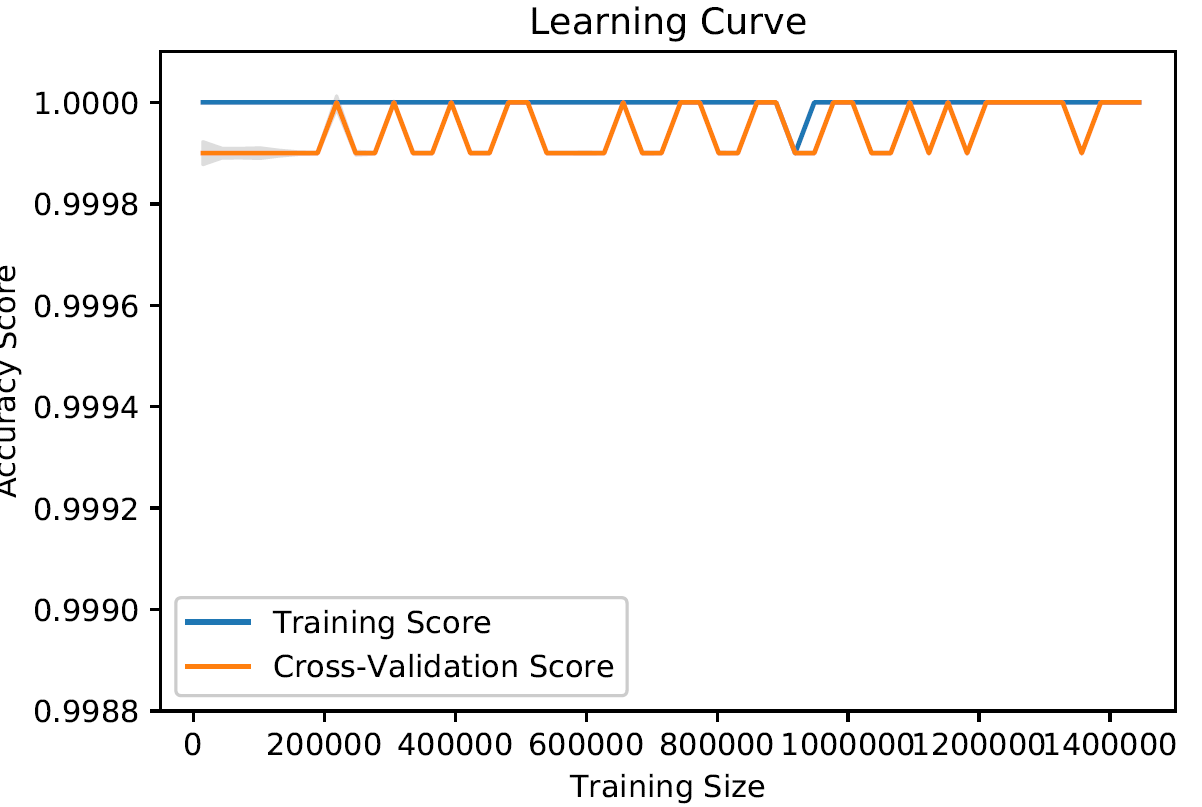}
	\caption{Learning Curve Showing Training and Cross-Validation Accuracy for UNSW-NB 2015 Dataset After CBFS }
	\label{unsw_correlation}
\end{figure}
\end{enumerate}
Similarly for the case of using CBFS method, it can be observed from Figs. \ref{cicids_correlation} and \ref{unsw_correlation} that the required training sample size for the CICIDS 2017 and UNSW-NB 2015 datasets is reduced to 500,000 and 200,000, respectively. This represents a reduction of 17\% and 75\% when compared to the required training sample size after SMOTE technique is applied. This shows that the CBFS method is also able to select relevant features that have a positive impact on the learning process. However, since some of the features selected may be redundant, this may have a negative impact on the learning process when compared to that of the IGBFS. The time needed to build the learning curve using the two feature selection methods was determined to further highlight the impact of feature selection on the reduction of time complexity. For example, building the learning curve for the UNSW-NB 2015 dataset required around 21 minutes and 25 minutes for the IGBFS and CBFS methods, respectively. Accordingly, applying either of the two feature selection methods will have a positive impact on the feature set size and training sample size with the IGBFS method having a slight advantage over the CBFS method.\\
\indent Figs. \ref{cicids_information}, \ref{unsw_information}, \ref{cicids_correlation}, and \ref{unsw_correlation} describe a relatively small or zero error gap between the training accuracy and the cross-validation accuracy. This indicates that the model is suitable to be generalized for testing and cross-validation datasets and is not being overfit to the training dataset \cite{overfitting_indicator}. 
\begin{table}[!tb]
	\caption{Optimal Parameter Values with IGBFS for Different ML Models }
	\centering%
	\scalebox{0.75}{
		\begin{tabular}{p{1.7cm}|p{3.5cm}|p{3.5cm}|}
			\toprule 
			& CICIDS 2017 &UNSW-NB 2015 \tabularnewline
			\cmidrule{2-3}  
			Classifier & Parameter Values & Parameter Values \tabularnewline
			\midrule
			%		RS-SVM &C = , gamma =  &C = 6.87, gamma = 6.61x10$^{-1}$  \tabularnewline
			%		PSO-SVM &C = , gamma =  &C = , gamma =  \tabularnewline
			%		GA-SVM &C = , gamma =  &C = 2.44, gamma = 7.49x10$^{-2}$ \tabularnewline
			%		BO-GP-SVM &C = 2, gamma =9.5x10$^-1$  & C = 3.42, gamma = 5.48x10$^{-5}$\tabularnewline
			%		BO-TPE-SVM &C = , gamma =  &C = 3.42, gamma = 5.48x10$^{-5}$\tabularnewline
			%		\bottomrule
			RS-KNN &Number of Neighbors= 3 &Number of Neighbors= 11  \tabularnewline
			PSO-KNN &Number of Neighbors= 5 &Number of Neighbors= 11  \tabularnewline
			GA-KNN &Number of Neighbors= 29 &Number of Neighbors= 13  \tabularnewline
			BO-GP-KNN&Number of Neighbors= 29 &Number of Neighbors= 13 \tabularnewline
			BO-TPE-KNN&Number of Neighbors= 29 &Number of Neighbors= 13  \tabularnewline
			\bottomrule
			RS-RF &Splitting Criterion= Gini, Number of trees= 40 &Splitting Criterion= Entropy, Number of trees= 30 \tabularnewline
			PSO-RF &Splitting Criterion= Gini, Number of trees= 21&Splitting Criterion= Entropy, Number of trees= 81\tabularnewline
			GA-RF &Splitting Criterion= Gini, Number of trees= 219 &Splitting Criterion= Entropy, Number of trees= 168\tabularnewline
			BO-GP-RF &Splitting Criterion= Entropy, Number of trees= 200 &Splitting Criterion= Entropy, Number of trees= 171\tabularnewline
			BO-TPE-RF &Splitting Criterion= Entropy, Number of trees= 90 &Splitting Criterion= Entropy, Number of trees= 50\tabularnewline
			\bottomrule
	\end{tabular}}
	\label{IGBFS_param}
\end{table}
\begin{table}[!tb]
	\caption{Optimal Parameter Values with CBFS for Different ML Models}
	\centering%
	\scalebox{0.75}{
		\begin{tabular}{p{1.7cm}|p{3.5cm}|p{3.5cm}|}
			\toprule 
			& CICIDS 2017 &UNSW-NB 2015 \tabularnewline
			\cmidrule{2-3}  
			Classifier & Parameter Values & Parameter Values \tabularnewline
			\midrule
			%		RS-SVM &C = , gamma =  &C = , gamma =   \tabularnewline
			%		PSO-SVM &C = , gamma =  &C = , gamma =  \tabularnewline
			%		GA-SVM &C = , gamma =  &C = 2.13, gamma = 4.21x10$^{-2}$ \tabularnewline
			%		BO-GP-SVM &C = , gamma =  & C = 8.42, gamma = 3.65x10$^{-4}$\tabularnewline
			%		BO-TPE-SVM &C = , gamma =  &C = 8.42, gamma = 3.65x10$^{-4}$\tabularnewline
			%		\bottomrule
			RS-KNN &Number of Neighbors= 3 &Number of Neighbors= 3  \tabularnewline
			PSO-KNN &Number of Neighbors= 11 &Number of Neighbors=  5 \tabularnewline
			GA-KNN &Number of Neighbors= 25 &Number of Neighbors=  25 \tabularnewline
			BO-GP-KNN&Number of Neighbors= 29 &Number of Neighbors= 25  \tabularnewline
			BO-TPE-KNN&Number of Neighbors= 29 &Number of Neighbors= 29  \tabularnewline
			\bottomrule
			RS-RF &Splitting Criterion= Gini, Number of trees= 20&Splitting Criterion= Gini, Number of trees= 10\tabularnewline
			PSO-RF &Splitting Criterion= Gini, Number of trees= 87&Splitting Criterion= Gini, Number of trees= 54 \tabularnewline
			GA-RF &Splitting Criterion= Gini, Number of trees= 219 &Splitting Criterion= Gini, Number of trees= 219\tabularnewline
			BO-GP-RF &Splitting Criterion= Gini, Number of trees= 164 &Splitting Criterion= Entropy, Number of trees= 78\tabularnewline
			BO-TPE-RF &Splitting Criterion= Entropy, Number of trees= 50 &Splitting Criterion= Gini, Number of trees= 20\tabularnewline
			\bottomrule
	\end{tabular}}
	\label{CBFS_param}
\end{table}
\begin{table*}[!tp]
	\caption{Performance Results of the Multi-stage Optimized ML-based NIDS Framework with IGBFS for Testing Datasets }
	\centering%
	\scalebox{0.85}{
		\begin{tabular}{l|cccc|cccc|}
			\toprule 
			& \multicolumn{4}{c|}{CICIDS 2017} & \multicolumn{4}{c|}{UNSW-NB 2015}\tabularnewline
			\cmidrule{2-5} \cmidrule{6-9} 
			Classifier & Acc(\%) & Precision & Recall & FAR & Acc(\%) & Precision & Recall & FAR\tabularnewline
			\midrule
			%		RS-SVM &  &  &  &   &100\%  &1.0  &1.0  &0.0 \tabularnewline
			%		PSO-SVM &  &  &   &  &100\%  &1.0  &1.0 &0.0 \tabularnewline
			%		GA-SVM &  &  &    &  &100\%  &1.0  &1.0 &0.0 \tabularnewline
			%		BO-GP-SVM &  &    &  & &100\%  &1.0  &1.0 &0.0 \tabularnewline
			%		BO-TPE-SVM &  &   &  &  &100\%  &1.0  &1.0 &0.0 \tabularnewline
			%		\bottomrule
			RS-KNN &99.63\%  &0.99  &0.99 &0.001  &99.96\%  &0.99  &0.99  &0.001 \tabularnewline
			PSO-KNN &99.09\%  &0.98  &0.99 &0.001  &99.91\%  &0.99  &0.99 &0.001 \tabularnewline
			GA-KNN &99.09\%  &0.98  &0.99 &0.001   &99.91\%  &0.99  &0.99 &0.001 \tabularnewline
			BO-GP-KNN &99.11\%  &0.98    &0.99  &0.001  &99.91\%  &0.99  &0.99 &0.001\tabularnewline
			BO-TPE-KNN &99.11\%  &0.98    &0.99  &0.001  &99.91\%  &0.99  &0.99 &0.001 \tabularnewline
			\bottomrule
			RS-RF &99.72\% &0.99 &0.99  &0.001  &100\%  &1.0  &1.0  &0.0\tabularnewline
			PSO-RF &99.98\%  &0.99 &0.99  &0.001  &100\%  &1.0  &1.0  &0.0 \tabularnewline
			GA-RF &99.98\%  &0.99 &0.99  &0.001  &100\%  &1.0  &1.0  &0.0 \tabularnewline
			BO-GP-RF &99.83\% &0.99 &0.99  &0.001  &100\%  &1.0  &1.0  &0.0 \tabularnewline
			BO-TPE-RF &99.99\% &0.99 &0.99  &0.001  &100\%  &1.0  &1.0  &0.0 \tabularnewline
			\bottomrule
	\end{tabular}}
	\label{IGBFS_resutls}
\end{table*}
\begin{table*}[!tp]
	\caption{Performance results of the Multi-stage Optimized ML-based NIDS Framework with CBFS for Testing Datasets }
	\centering%
	\scalebox{0.85}{
		\begin{tabular}{l|cccc|cccc|}
			\toprule 
			& \multicolumn{4}{c|}{CICIDS 2017} & \multicolumn{4}{c|}{UNSW-NB 2015}\tabularnewline
			\cmidrule{2-5} \cmidrule{6-9} 
			Classifier & Acc(\%) & Precision & Recall & FAR & Acc(\%) & Precision & Recall & FAR\tabularnewline
			\midrule
			%		RS-SVM &  &  &  &   &  &  &  & \tabularnewline
			%		PSO-SVM &  &  &   &  &  &  &  & \tabularnewline
			%		GA-SVM &  &  &    &  &100\%  &1.0  &1.0 &0.0 \tabularnewline
			%		BO-GP-SVM &  &    &  & &100\%  &1.0  &1.0 &0.0 \tabularnewline
			%		BO-TPE-SVM &  &   &  &  &100\%  &1.0  &1.0 &0.0 \tabularnewline
			%		\bottomrule
			RS-KNN &99.70\%  &0.99 &0.99 &0.001  &99.96\% &0.99&0.99  &0.001 \tabularnewline
			PSO-KNN &99.28\%  &0.99  &0.99 &0.001  &99.88\%&0.99 &0.99  &0.001 \tabularnewline
			GA-KNN &99.28\% &0.99  & 0.99 &0.001 &99.88\%&0.99 &0.99  &0.001 \tabularnewline
			BO-GP-KNN &99.23\%  &0.99 &0.99 &0.001&99.88\%&0.99 &0.99  &0.001 \tabularnewline
			BO-TPE-KNN &99.23\%  &0.99  &0.99  &0.001  &99.88\%&0.99 &0.99  &0.001 \tabularnewline
			\bottomrule
			RS-RF &99.61\%  &0.99 & 0.99 &0.001  &100\%  &1.0  &1.0 &0.0 \tabularnewline
			PSO-RF&99.88\% &0.99 &0.99  &0.001  &100\%  &1.0  &1.0 &0.0 \tabularnewline
			GA-RF &99.88\% &0.99 &0.99  &0.001  &100\%  &1.0  &1.0 &0.0 \tabularnewline
			BO-GP-RF &99.88\% &0.99  &0.99  &0.001  &100\%  &1.0  &1.0 &0.0 \tabularnewline
			BO-TPE-RF &99.99\% &0.99 &0.99 &0.001  &100\%  &1.0  &1.0 &0.0 \tabularnewline
			\bottomrule
	\end{tabular}}
	\label{CBFS_resutls}
\end{table*}
\subsubsection{Impact of optimization methods on the ML models' detection performance}\mbox{}\\
\indent To evaluate the performance of the different classifiers and study the impact of the different optimization methods on them, we determine four evaluation metrics, namely the accuracy (acc), precision, recall/true positive rate (TPR), and false alarm/positive rate (FAR/FPR) as per \cite{sc}\cite{regression1} using the following equations:
\begin{equation}
Acc=\frac{tp+tn}{tp+tn+fp+fn}
\end{equation}
\begin{equation}
Precision=\frac{tp}{tp+fp}
\end{equation}
\begin{equation}
Recall/TPR=\frac{tp}{tp+fn}
\end{equation}
\begin{equation}
FAR/FPR=\frac{fp}{tn+fp}
\end{equation}
where $tp$ is the number of true positives, $tn$ is the number of true negatives, $fp$ is the number of false positives, and $fn$ is the number of false negatives. These values compose the confusion matrix of any ML model.\\
\indent Table \ref{IGBFS_param} gives the optimal parameter values for the two different classifiers when the IGBFS technique is used. In the case of KNN method, the RS and PSO methods tend to choose smaller values for the number of neighbors when compared to the GA, BO-GP, and BO-TPE methods. For the RS method, this is due to the fact that the algorithm's stopping criterion is typically the number of iterations and thereby does not test all potential values. Accordingly, it is possible for it to miss the optimal number of neighbors. Similarly, one of the stopping criteria in the PSO algorithm is also the number of evaluations, which can also lead to it missing the optimal value. In contrast, the GA, BO-GP, and BO-TPE all resulted in a similar number of neighbors for both the CICIDS 2017 and UNSW-NB 2015 datasets. For the GA algorithm, the number of generations is typically set sufficiently high to reach the optimal value for the number of neighbors. In a similar manner, the BO-GP and BO-TPE determine the actual optimal value based on the assumed model.\\
\indent In the case of the RF method, the RS and PSO algorithms tend to choose a lower number of trees compared to the GA, BO-GP, and BO-TPE. This is due to the algorithms' stopping criterion that often leads to a pre-mature stoppage. In contrast, the GA, BO-GP, and BO-TPE determine that the number of trees needed is higher as they explore more potential values, allowing them to select more optimal values for the number of trees. In terms of the splitting criterion, the entropy criterion is mostly selected. This is expected since the IGBFS method selects features based on their information gain, which is determined using the entropy of each feature. As such, this criterion would be more suitable when using IGBFS.\\
\indent Looking at Table \ref{CBFS_param}, similar observations about the hyper-parameter optimization performance of the different algorithms can be made for both the KNN and RF methods. The only difference is that for the RF method, the splitting criterion is chosen to be the Gini index. This is due to the CBFS method using the correlation as the selection criterion rather than the entropy. Therefore, the features chosen may result in a low amount of information (equivalent to having a high entropy with respect to the class), and thus would be overlooked if the entropy splitting criterion is chosen. This is the reason behind choosing the Gini splitting criterion when the CBFS method is used.\\
\indent Tables \ref{IGBFS_resutls} and \ref{CBFS_resutls} show the performance of the two classification algorithms when using IGBFS and CBFS methods, respectively. Several observations can be made. The first observation is that the optimized models outperform the regular models recently reported in \cite{cicids17}\cite{sd}\cite{unsw2015-2} by 1-2\% on average in terms of accuracy and a reduction of 1-2\% in FAR for both datasets. This is expected since one of the main goals of hyper-parameter optimization is to improve the performance of the ML models. The second observation is that the RF classifier outperforms the KNN classifier for both the IGBFS and CBFS methods as seen in the CICIDS 2017 and UNSW-NB 2015 datasets. This reiterates the previously obtained results in \cite{sc} with ISCX 2012 dataset and the reported results in \cite{cicids17}\cite{sd}\cite{unsw2015-2} in which the RF classifier also outperformed the KNN model. This can be attributed to the RF classifier being an ensemble model. Accordingly, it is effective with non-linear and high-dimensional datasets like the datasets under consideration in this work. The third observation is that the BO-TPE-RF method had the highest detection accuracy for both the CICIDS 2017 and UNSW-NB 2015 datasets for both feature selection algorithms with a detection accuracy of 99.99\% and 100\%, respectively. This proves the effectiveness and robustness of the proposed multi-stage optimized ML-based NIDS framework as it outperformed other NIDS frameworks.
\section{Conclusion}\label{conc}
\indent The area of cyber-security has garnered significant attention from both the industry and academia due to the increased dependency of individuals and organizations on the Internet and their concern about the security and privacy of their activities. More resources are being deployed and allocated to protect modern Internet-based networks against potential attacks or anomalous activities. Accordingly, different types of network intrusion detection systems (NIDSs) have been proposed in the literature. Despite the continuous improvements in NIDS performance, there is still room for further improvement. More insights can be extracted from the high volume of network traffic data generated, the continuously changing environments, the plethora of features collected as part of training datasets (high dimensional datasets), and the need for real-time intrusion detection.\\
\indent Choosing the most suitable subset of features and optimizing the parameters of the machine learning (ML)-based detection models is essential to enhance their performance. Accordingly, this paper expanded on our previous work by proposing a multi-stage optimized ML-based NIDS framework that reduced the computational complexity while maintaining its detection performance. Using two recent state-of-the-art intrusion detection datasets (CICIDS 2017 dataset and the UNSW-NB 2015 dataset) for performance evaluation, this work first studied the impact of oversampling techniques on the models' training sample size and determined the minimum suitable training size for effective intrusion detection. Experimental results showed that using the SMOTE oversampling technique can reduce the training sample size between 39\% and 74\% of the original datasets' size. Additionally, this work compared between two different feature selection techniques, namely information gain (IGBFS) and correlation-based feature selection (CBFS), and explored their impact on the feature set size, the training sample size, and the models' detection performance. The experimental results showed that the feature selection methods were able to reduce the feature set size by almost 60\%. Moreover, they further reduced the required training sample size between 33\% and 50\% when compared to the training sample after SMOTE. Finally, this work investigated the impact of different ML hyper-parameter optimization techniques on the NIDS's performance using two ML classification models, namely the K-nearest neighbors (KNN) and the Random Forest (RF) classifiers. Experimental results showed that the optimized RF classifier with Bayesian Optimization using Tree Parzen Estimator (BO-TPE-RF) had the highest detection accuracy when compared to the other optimization techniques (enhanced the detection accuracy by 1-2\% and reduce the FAR by 1-2\% when compared to recent works from the literature). It was also observed that using the IGBFS method achieved better detection accuracy when compared to the CBFS method.\\
\indent Other models such as deep learning classifiers can be explored for future work since these models perform admirably on non-linear and high-dimensional datasets. Investigating the impact of combining supervised and unsupervised ML techniques may also prove paramount in this field to detect novel attacks. 

%\balance
\bibliographystyle{IEEEtran}
\bibliography{bibfile}

\begin{thebibliography}{10}
\providecommand{\url}[1]{#1}
\csname url@samestyle\endcsname
\providecommand{\newblock}{\relax}
\providecommand{\bibinfo}[2]{#2}
\providecommand{\BIBentrySTDinterwordspacing}{\spaceskip=0pt\relax}
\providecommand{\BIBentryALTinterwordstretchfactor}{4}
\providecommand{\BIBentryALTinterwordspacing}{\spaceskip=\fontdimen2\font plus
\BIBentryALTinterwordstretchfactor\fontdimen3\font minus
  \fontdimen4\font\relax}
\providecommand{\BIBforeignlanguage}[2]{{%
\expandafter\ifx\csname l@#1\endcsname\relax
\typeout{** WARNING: IEEEtran.bst: No hyphenation pattern has been}%
\typeout{** loaded for the language `#1'. Using the pattern for}%
\typeout{** the default language instead.}%
\else
\language=\csname l@#1\endcsname
\fi
#2}}
\providecommand{\BIBdecl}{\relax}
\BIBdecl

\bibitem{bib1}
C.-F. Tsai, Y.-F. Hsu, C.-Y. Lin, and W.-Y. Lin, ``Intrusion detection by
  machine learning: A review,'' \emph{Expert Systems with Applications},
  vol.~36, no.~10, pp. 11\,994--12\,000, 2009.

\bibitem{sb}
A.~Moubayed, M.~Injadat, A.~Shami, and H.~Lutfiyya, ``Student engagement level
  in e learning environment: Clustering using k-means,'' \emph{American Journal
  of Distance Education}, vol.~34, no.~2, 2019.

\bibitem{bib25}
\BIBentryALTinterwordspacing
M.~Injadat, F.~Salo, and A.~B. Nassif, ``Data mining techniques in social
  media: A survey,'' \emph{Neurocomputing}, vol. 214, pp. 654 -- 670, 2016.
  [Online]. Available:
  \url{http://www.sciencedirect.com/science/article/pii/S092523121630683X}
\BIBentrySTDinterwordspacing

\bibitem{bib2}
M.~B. Salem, S.~Hershkop, and S.~J. Stolfo, ``A survey of insider attack
  detection research,'' in \emph{Insider Attack and Cyber Security}.\hskip 1em
  plus 0.5em minus 0.4em\relax Springer, 2008, pp. 69--90.

\bibitem{bib3}
W.~Bul'ajoul, A.~James, and M.~Pannu, ``Improving network intrusion detection
  system performance through quality of service configuration and parallel
  technology,'' \emph{Journal of Computer and System Sciences}, vol.~81, no.~6,
  pp. 981--999, 2015.

\bibitem{bib4}
S.~M.~H. Bamakan, B.~Amiri, M.~Mirzabagheri, and Y.~Shi, ``A new intrusion
  detection approach using pso based multiple criteria linear programming,''
  \emph{Procedia Computer Science}, vol.~55, pp. 231--237, 2015.

\bibitem{bib5}
S.~X. Wu and W.~Banzhaf, ``The use of computational intelligence in intrusion
  detection systems: A review,'' \emph{Applied soft computing}, vol.~10, no.~1,
  pp. 1--35, 2010.

\bibitem{bib6}
H.-J. Liao, C.-H.~R. Lin, Y.-C. Lin, and K.-Y. Tung, ``Intrusion detection
  system: A comprehensive review,'' \emph{Journal of Network and Computer
  Applications}, vol.~36, no.~1, pp. 16--24, 2013.

\bibitem{bib7}
S.~Suthaharan, ``Big data classification: Problems and challenges in network
  intrusion prediction with machine learning,'' \emph{ACM SIGMETRICS
  Performance Evaluation Review}, vol.~41, no.~4, pp. 70--73, 2014.

\bibitem{bib8}
J.~Zhang and M.~Zulkernine, ``Anomaly based network intrusion detection with
  unsupervised outlier detection,'' in \emph{Communications, 2006. ICC'06. IEEE
  International Conference on}, vol.~5.\hskip 1em plus 0.5em minus 0.4em\relax
  IEEE, 2006, pp. 2388--2393.

\bibitem{sc}
M.~{Injadat}, F.~{Salo}, A.~B. {Nassif}, A.~{Essex}, and A.~{Shami}, ``Bayesian
  optimization with machine learning algorithms towards anomaly detection,'' in
  \emph{2018 IEEE Global Communications Conference (GLOBECOM)}, 2018, pp. 1--6.

\bibitem{cicids17}
I.~Sharafaldin, A.~H. Lashkari, and A.~A. Ghorbani, ``Toward generating a new
  intrusion detection dataset and intrusion traffic characterization.'' in
  \emph{ICISSP}, 2018, pp. 108--116.

\bibitem{bib9}
A.~Shiravi, H.~Shiravi, M.~Tavallaee, and A.~A. Ghorbani, ``Toward developing a
  systematic approach to generate benchmark datasets for intrusion detection,''
  \emph{Computers and Security}, vol.~31, no.~3, pp. 357 -- 374, 2012.

\bibitem{unsw15}
N.~{Moustafa} and J.~{Slay}, ``Unsw-nb15: a comprehensive data set for network
  intrusion detection systems (unsw-nb15 network data set),'' in \emph{2015
  Military Communications and Information Systems Conference (MilCIS)}, Nov.
  2015, pp. 1--6.

\bibitem{bib10}
F.~Kuang, W.~Xu, and S.~Zhang, ``A novel hybrid kpca and svm with ga model for
  intrusion detection,'' \emph{Applied Soft Computing}, vol.~18, pp. 178--184,
  2014.

\bibitem{bib11}
A.~S. Eesa, Z.~Orman, and A.~M.~A. Brifcani, ``A novel feature-selection
  approach based on the cuttlefish optimization algorithm for intrusion
  detection systems,'' \emph{Expert Systems with Applications}, vol.~42, no.~5,
  pp. 2670--2679, 2015.

\bibitem{bib12}
W.~Li, P.~Yi, Y.~Wu, L.~Pan, and J.~Li, ``A new intrusion detection system
  based on knn classification algorithm in wireless sensor network,''
  \emph{Journal of Electrical and Computer Engineering}, vol. 2014, 2014.

\bibitem{ann1}
A.~B. {Nassif}, L.~F. {Capretz}, and D.~{Ho}, ``Estimating software effort
  using an ann model based on use case points,'' in \emph{2012 11th
  International Conference on Machine Learning and Applications}, vol.~2, 2012,
  pp. 42--47.

\bibitem{s11}
A.~{Moubayed}, M.~{Injadat}, A.~B. {Nassif}, H.~{Lutfiyya}, and A.~{Shami},
  ``E-learning: Challenges and research opportunities using machine learning
  data analytics,'' \emph{IEEE Access}, vol.~6, pp. 39\,117--39\,138, 2018.

\bibitem{bib13}
S.~Aljawarneh, M.~Aldwairi, and M.~B. Yassein, ``Anomaly-based intrusion
  detection system through feature selection analysis and building hybrid
  efficient model,'' \emph{Journal of Computational Science}, 2017.

\bibitem{se}
A.~Moubayed, M.~Injadat, A.~Shami, and H.~Lutfiyya, ``Dns typo-squatting domain
  detection: A data analytics and machine learning based approach,'' in
  \emph{2018 IEEE Global Communications Conference (GLOBECOM)}.\hskip 1em plus
  0.5em minus 0.4em\relax IEEE, 2018, pp. 1--7.

\bibitem{se2}
A.~Moubayed, E.~Aqeeli, and A.~Shami, ``{Ensemble-based Feature Selection and
  Classification Model for DNS Typo-squatting Detection},'' in \emph{{Accepted
  in 33rd Canadian Conference on Electrical and Computer Engineering
  (CCECE'20}}.\hskip 1em plus 0.5em minus 0.4em\relax IEEE, 2020, pp. 1--6.

\bibitem{forensics1}
X.~{Sun}, J.~{Dai}, P.~{Liu}, A.~{Singhal}, and J.~{Yen}, ``Using bayesian
  networks for probabilistic identification of zero-day attack paths,''
  \emph{IEEE Transactions on Information Forensics and Security}, vol.~13,
  no.~10, pp. 2506--2521, Oct. 2018.

\bibitem{TNSM2}
A.~{Alsirhani}, S.~{Sampalli}, and P.~{Bodorik}, ``Ddos detection system: Using
  a set of classification algorithms controlled by fuzzy logic system in apache
  spark,'' \emph{IEEE Transactions on Network and Service Management}, vol.~16,
  no.~3, pp. 936--949, Sep. 2019.

\bibitem{TNSM1}
A.~A. {Daya}, M.~A. {Salahuddin}, N.~{Limam}, and R.~{Boutaba}, ``Botchase:
  Graph-based bot detection using machine learning,'' \emph{IEEE Transactions
  on Network and Service Management}, pp. 1--1, 2020.

\bibitem{forensics2}
T.~{Kim}, B.~{Kang}, M.~{Rho}, S.~{Sezer}, and E.~G. {Im}, ``A multimodal deep
  learning method for android malware detection using various features,''
  \emph{IEEE Transactions on Information Forensics and Security}, vol.~14,
  no.~3, pp. 773--788, Mar. 2019.

\bibitem{sf}
F.~Salo, M.~Injadat, A.~B. Nassif, A.~Shami, and A.~Essex, ``Data mining
  techniques in intrusion detection systems: A systematic literature review,''
  \emph{IEEE Access}, vol.~6, pp. 56\,046--56\,058, 2018.

\bibitem{sf1}
F.~Salo, M.~Injadat, A.~B. Nassif, and A.~Essex, ``Data mining with big data in
  intrusion detection systems: A systematic literature review,'' in
  \emph{International Symposium on Big Data Management and Analytics 2019},
  Apr. 2019.

\bibitem{sg}
F.~Salo, M.~Injadat, A.~Moubayed, A.~B. Nassif, and A.~Essex, ``Clustering
  enabled classification using ensemble feature selection for intrusion
  detection,'' in \emph{2019 International Conference on Computing, Networking
  and Communications (ICNC)}.\hskip 1em plus 0.5em minus 0.4em\relax IEEE,
  2019, pp. 276--281.

\bibitem{sd}
L.~Yang, A.~Moubayed, I.~Hamieh, and A.~Shami, ``Tree-based intelligent
  intrusion detection system in internet of vehicles,'' in \emph{2019 IEEE
  Global Communications Conference (GLOBECOM)}, 2019.

\bibitem{bib14}
Y.~Y. Chung and N.~Wahid, ``A hybrid network intrusion detection system using
  simplified swarm optimization (sso),'' \emph{Applied Soft Computing},
  vol.~12, no.~9, pp. 3014--3022, 2012.

\bibitem{bib15}
F.~Kuang, W.~Xu, and S.~Zhang, ``A novel hybrid kpca and svm with ga model for
  intrusion detection,'' \emph{Applied Soft Computing}, vol.~18, pp. 178--184,
  2014.

\bibitem{bib17}
J.~Zhang, M.~Zulkernine, and A.~Haque, ``Random-forests-based network intrusion
  detection systems,'' \emph{IEEE Transactions on Systems, Man, and
  Cybernetics, Part C (Applications and Reviews)}, vol.~38, no.~5, pp.
  649--659, 2008.

\bibitem{zscore}
K.~M. Ali~Alheeti and K.~McDonald-Maier, ``Intelligent intrusion detection in
  external communication systems for autonomous vehicles,'' \emph{Systems
  Science and Control Engineering}, vol.~6, no.~1, pp. 48--56, 2018.

\bibitem{traffic_imbalance}
Z.~Chen, Q.~Yan, H.~Han, S.~Wang, L.~Peng, L.~Wang, and B.~Yang, ``Machine
  learning based mobile malware detection using highly imbalanced network
  traffic,'' \emph{Information Sciences}, vol. 433, pp. 346--364, 2018.

\bibitem{SMOTE}
N.~V. Chawla, K.~W. Bowyer, L.~O. Hall, and W.~P. Kegelmeyer, ``Smote:
  synthetic minority over-sampling technique,'' \emph{Journal of artificial
  intelligence research}, vol.~16, pp. 321--357, 2002.

\bibitem{SMOTE2}
X.~Tan, S.~Su, Z.~Huang, X.~Guo, Z.~Zuo, X.~Sun, and L.~Li, ``Wireless sensor
  networks intrusion detection based on smote and the random forest
  algorithm,'' \emph{Sensors}, vol.~19, no.~1, p. 203, 2019.

\bibitem{FS_reason}
M.~B. {Çatalkaya}, O.~{Kalıpsız}, M.~S. {Aktaş}, and U.~O. {Turgut}, ``Data
  feature selection methods on distributed big data processing platforms,'' in
  \emph{2018 3rd International Conference on Computer Science and Engineering
  (UBMK)}, Sep. 2018, pp. 133--138.

\bibitem{FS_IG1}
R.~S.~B. {Krishna} and M.~{Aramudhan}, ``{Feature Selection Based on
  Information Theory for Pattern Classification},'' in \emph{2014 International
  Conference on Control, Instrumentation, Communication and Computational
  Technologies (ICCICCT)}, Jul. 2014, pp. 1233--1236.

\bibitem{FS_IG2}
B.~Bonev, ``Feature selection based on information theory,'' Ph.D.
  dissertation, University of Alicante, Jun. 2010.

\bibitem{FS_CFS1}
J.~Li, K.~Cheng, S.~Wang, F.~Morstatter, R.~P. Trevino, J.~Tang, and H.~Liu,
  ``Feature selection: A data perspective,'' \emph{ACM Computing Surveys
  (CSUR)}, vol.~50, no.~6, p.~94, 2018.

\bibitem{FS_CFS2}
M.~A. Hall, ``Correlation-based feature selection for machine learning,'' Ph.D.
  dissertation, University of Waikato Hamilton, 1999.

\bibitem{sa}
A.~{Moubayed}, M.~{Injadat}, A.~{Shami}, and H.~{Lutfiyya}, ``Relationship
  between student engagement and performance in e learning environment using
  association rules,'' in \emph{2018 IEEE World Engineering Education
  Conference (EDUNINE)}, Mar. 2018, pp. 1--6.

\bibitem{FS_CFS3}
J.~H. Gennari, P.~Langley, and D.~Fisher, ``Models of incremental concept
  formation,'' \emph{Artificial intelligence}, vol.~40, no. 1-3, pp. 11--61,
  1989.

\bibitem{ee}
P.~Koch, B.~Wujek, O.~Golovidov, and S.~Gardner, ``Automated hyperparameter
  tuning for effective machine learning,'' in \emph{Proceedings of the SAS
  Global Forum 2017 Conference}, 2017, pp. 1--23.

\bibitem{hyper1}
\BIBentryALTinterwordspacing
L.~Yang and A.~Shami, ``On hyperparameter optimization of machine learning
  algorithms: Theory and practice,'' \emph{Neurocomputing}, 2020. [Online].
  Available:
  \url{http://www.sciencedirect.com/science/article/pii/S0925231220311693}
\BIBentrySTDinterwordspacing

\bibitem{RS_heuristic}
J.~Bergstra and Y.~Bengio, ``Random search for hyper-parameter optimization,''
  \emph{Journal of machine learning research}, vol.~13, no. Feb, pp. 281--305,
  2012.

\bibitem{injadat_ch4}
\BIBentryALTinterwordspacing
M.~Injadat, A.~Moubayed, A.~B. Nassif, and A.~Shami, ``{Systematic ensemble
  model selection approach for educational data mining},''
  \emph{Knowledge-Based Systems}, vol. 200, p. 105992, 2020. [Online].
  Available:
  \url{http://www.sciencedirect.com/science/article/pii/S0950705120302999}
\BIBentrySTDinterwordspacing

\bibitem{injadat_ch5}
------, ``{Multi-split Optimized Bagging Ensemble Model Selection for
  Multi-class Educational Datasets},'' \emph{Applied Intelligence}, 2020.

\bibitem{metaheuristic}
L.~Bianchi, M.~Dorigo, L.~M. Gambardella, and W.~J. Gutjahr, ``A survey on
  metaheuristics for stochastic combinatorial optimization,'' \emph{Natural
  Computing}, vol.~8, no.~2, pp. 239--287, 2009.

\bibitem{PSO1}
S.-W. Lin, K.-C. Ying, S.-C. Chen, and Z.-J. Lee, ``Particle swarm optimization
  for parameter determination and feature selection of support vector
  machines,'' \emph{Expert Systems with Applications}, vol.~35, no.~4, pp. 1817
  -- 1824, 2008.

\bibitem{GA1}
G.~Cohen, M.~Hilario, and A.~Geissbuhler, ``Model selection for support vector
  classifiers via genetic algorithms. an application to medical decision
  support,'' in \emph{International Symposium on Biological and Medical Data
  Analysis}.\hskip 1em plus 0.5em minus 0.4em\relax Springer, 2004, pp.
  200--211.

\bibitem{GA_app1}
S.~G. Ahmad, C.~S. Liew, E.~U. Munir, T.~F. Ang, and S.~U. Khan, ``A hybrid
  genetic algorithm for optimization of scheduling workflow applications in
  heterogeneous computing systems,'' \emph{Journal of Parallel and Distributed
  Computing}, vol.~87, pp. 80--90, 2016.

\bibitem{GA_app2}
S.~Blaifi, S.~Moulahoum, I.~Colak, and W.~Merrouche, ``An enhanced dynamic
  model of battery using genetic algorithm suitable for photovoltaic
  applications,'' \emph{Applied Energy}, vol. 169, pp. 888--898, 2016.

\bibitem{GA_app3}
U.~Mehboob, J.~Qadir, S.~Ali, and A.~Vasilakos, ``Genetic algorithms in
  wireless networking: techniques, applications, and issues,'' \emph{Soft
  Computing}, vol.~20, no.~6, pp. 2467--2501, 2016.

\bibitem{GA_app4}
A.~Rikhtegar, M.~Pooyan, and M.~T. Manzuri-Shalmani, ``Genetic
  algorithm-optimised structure of convolutional neural network for face
  recognition applications,'' \emph{IET Computer Vision}, vol.~10, no.~6, pp.
  559--566, 2016.

\bibitem{BO1}
J.~Snoek, H.~Larochelle, and R.~P. Adams, ``Practical bayesian optimization of
  machine learning algorithms,'' in \emph{Advances in neural information
  processing systems}, 2012, pp. 2951--2959.

\bibitem{bib23}
E.~Brochu, V.~M. Cora, and N.~De~Freitas, ``A tutorial on bayesian optimization
  of expensive cost functions, with application to active user modeling and
  hierarchical reinforcement learning,'' \emph{arXiv preprint arXiv:1012.2599},
  2010.

\bibitem{BO2}
I.~Dewancker, M.~McCourt, and S.~Clark, ``Bayesian optimization primer,'' 2015.

\bibitem{BO3}
K.~Eggensperger, M.~Feurer, F.~Hutter, J.~Bergstra, J.~Snoek, H.~Hoos, and
  K.~Leyton-Brown, ``Towards an empirical foundation for assessing bayesian
  optimization of hyperparameters,'' in \emph{NIPS workshop on Bayesian
  Optimization in Theory and Practice}, vol.~10, 2013, p.~3.

\bibitem{evolutionary1}
D.~Ashlock, \emph{Evolutionary computation for modeling and
  optimization}.\hskip 1em plus 0.5em minus 0.4em\relax Springer Science and
  Business Media, 2006.

\bibitem{sdp1}
A.~{Moubayed}, A.~{Refaey}, and A.~{Shami}, ``Software-defined perimeter (sdp):
  State of the art secure solution for modern networks,'' \emph{IEEE Network},
  vol.~33, no.~5, pp. 226--233, 2019.

\bibitem{sdp2}
P.~{Kumar}, A.~{Moubayed}, A.~{Refaey}, A.~{Shami}, and J.~{Koilpillai},
  ``Performance analysis of sdp for secure internal enterprises,'' in
  \emph{2019 IEEE Wireless Communications and Networking Conference (WCNC)},
  2019, pp. 1--6.

\bibitem{SMOTE_complexity}
F.~Hu and H.~Li, ``A novel boundary oversampling algorithm based on
  neighborhood rough set model: Nrsboundary-smote,'' \emph{Mathematical
  Problems in Engineering}, vol. 2013, 2013.

\bibitem{RS_complexity}
A.~Lissovoi, P.~S. Oliveto, and J.~A. Warwicker, ``On the time complexity of
  algorithm selection hyper-heuristics for multimodal optimisation,'' in
  \emph{Proceedings of the AAAI Conference on Artificial Intelligence},
  vol.~33, 2019, pp. 2322--2329.

\bibitem{PSO_complexity}
R.~Cheng and Y.~Jin, ``A social learning particle swarm optimization algorithm
  for scalable optimization,'' \emph{Information Sciences}, vol. 291, pp.
  43--60, 2015.

\bibitem{GA_complexity}
P.~S. Oliveto and C.~Witt, ``Improved time complexity analysis of the simple
  genetic algorithm,'' \emph{Theoretical Computer Science}, vol. 605, pp.
  21--41, 2015.

\bibitem{BO_TPE_complexity}
M.~Feurer and F.~Hutter, ``Hyperparameter optimization,'' in \emph{Automated
  Machine Learning}.\hskip 1em plus 0.5em minus 0.4em\relax Springer, 2019, pp.
  3--33.

\bibitem{complexity1}
{The Kernel Trip}, ``{Computational complexity of machine learning
  algorithms},'' Apr. 2018.

\bibitem{complexity2}
C.-T. Chu, S.~K. Kim, Y.-A. Lin, Y.~Yu, G.~Bradski, K.~Olukotun, and A.~Y. Ng,
  ``Map-reduce for machine learning on multicore,'' in \emph{Advances in neural
  information processing systems}, 2007, pp. 281--288.

\bibitem{svm_complexity}
F.~Pedregosa, G.~Varoquaux, A.~Gramfort, V.~Michel, B.~Thirion, O.~Grisel,
  M.~Blondel, P.~Prettenhofer, R.~Weiss, V.~Dubourg, J.~Vanderplas, A.~Passos,
  D.~Cournapeau, M.~Brucher, M.~Perrot, and E.~Duchesnay, ``Scikit-learn:
  Machine learning in {P}ython,'' \emph{Journal of Machine Learning Research},
  vol.~12, pp. 2825--2830, 2011.

\bibitem{KNN_testing_complexity}
\BIBentryALTinterwordspacing
O.~Veksler, ``Cs434a/541a class notes prof. olga veksler,'' 2015. [Online].
  Available: \url{http://www.csd.uwo.ca/courses/CS9840a/Lecture2_knn.pdf}
\BIBentrySTDinterwordspacing

\bibitem{RF_testing_complexity}
X.~Sol{\'e}, A.~Ramisa, and C.~Torras, ``Evaluation of random forests on
  large-scale classification problems using a bag-of-visual-words
  representation.'' in \emph{CCIA}, 2014, pp. 273--276.

\bibitem{overfitting_indicator}
G.~James, D.~Witten, T.~Hastie, and R.~Tibshirani, \emph{An introduction to
  statistical learning}.\hskip 1em plus 0.5em minus 0.4em\relax Springer, 2013,
  vol. 112.

\bibitem{regression1}
A.~B. Nassif, D.~Ho, and L.~F. Capretz, ``Regression model for software effort
  estimation based on the use case point method,'' in \emph{2011 International
  Conference on Computer and Software Modeling}, vol.~14, 2011, pp. 106--110.

\bibitem{unsw2015-2}
N.~{Moustafa}, B.~{Turnbull}, and K.~R. {Choo}, ``An ensemble intrusion
  detection technique based on proposed statistical flow features for
  protecting network traffic of internet of things,'' \emph{IEEE Internet of
  Things Journal}, vol.~6, no.~3, pp. 4815--4830, 2019.

\end{thebibliography}
\vspace{-1.25cm}
\begin{IEEEbiography}[{\includegraphics[width=1in,height=1.25in,clip,keepaspectratio]{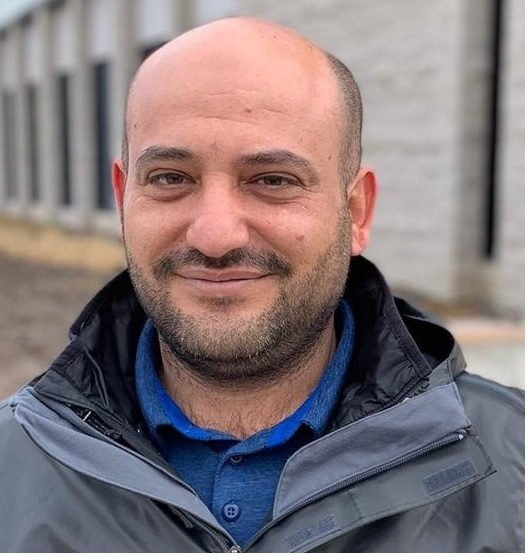}}]{MohammadNoor Injadat} received the B.Sc. and M.Sc. degrees in Computer Science from Al Al-Bayt University and University Putra Malaysia in Jordan and Malaysia in 2000 and 2002, respectively. He obtained a Master of Engineering in Electrical and Computer Engineering from University of Western Ontario in 2015. He obtained his Ph.D. degree in Software Engineering at the Department of Electrical and Computer Engineering, University of Western Ontario in Canada in 2020. His research interests include data mining, machine learning, social network analysis, e-learning analytics, and network security.
\end{IEEEbiography}
\vspace{-1.25cm}
\begin{IEEEbiography}[{\includegraphics[width=1in,height=1.25in,clip,keepaspectratio]{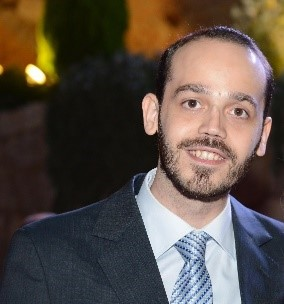}}]{Abdallah Moubayed} received his Ph.D. in Electrical \& Computer Engineering from the University of Western Ontario in August 2018, his M.Sc. degree in Electrical Engineering from King Abdullah University of Science and Technology, Thuwal, Saudi Arabia in 2014, and his B.E. degree in Electrical Engineering from the Lebanese American University, Beirut, Lebanon in 2012. Currently, he is a Postdoctoral Associate in the Optimized Computing and Communications (OC2) lab at University of Western Ontario. His research interests include wireless communication, resource allocation, wireless network virtualization, performance \& optimization modeling, machine learning \& data analytics, computer network security, cloud computing, and e-learning.
\end{IEEEbiography}
\vspace{-1.25cm}
\begin{IEEEbiography}[{\includegraphics[width=1in,height=1.25in,clip,keepaspectratio]{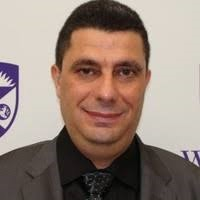}}]{Ali Bou Nassif} received his Ph.D. degree in Electrical and Computer engineering from The University of Western Ontario, London, Ontario, Canada (2012). He is currently an Assistant Professor and an Assistant Dean of the Graduate Studies, University of Sharjah, United Arab Emirates, and an Adjunct Research Professor with Western University. He has published more than 60 papers in international journals and conferences. His interests are Machine Learning and Soft Computing, Software Engineering, Cloud Computing and Service Oriented Architecture (SOA), and Mobile Computing. Ali is a registered professional engineer in Ontario, as well as a member of IEEE Computer Society.
\end{IEEEbiography}
\vspace{-1.25cm}
\begin{IEEEbiography}[{\includegraphics[width=1in,height=1.25in,clip,keepaspectratio]{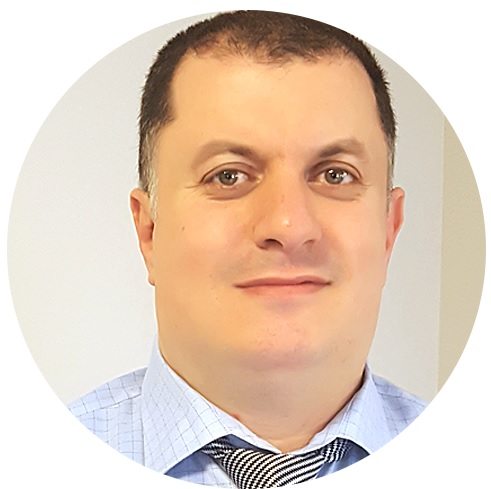}}]{Abdallah Shami} is a Professor at the ECE department at Western University, Ontario, Canada. Dr. Shami is the Director of the Optimized Computing and Communications Laboratory at Western. He is currently an Associate Editor for IEEE Transactions on Mobile Computing, IEEE Network, and IEEE Communications Tutorials and Survey. Dr. Shami has chaired key symposia for IEEE GLOBECOM, IEEE ICC, IEEE ICNC, and ICCIT. He was the elected Chair of the IEEE Communications Society Technical Committee on Communications Software (2016-2017) and IEEE London Ontario Section Chair (2016-2018).
\end{IEEEbiography}

\end{document}